\documentclass[times,twocolumn,final]{elsarticle} 
\usepackage{medima}
\usepackage{framed,multirow}

\usepackage{latexsym}

\usepackage{url}
\usepackage{xcolor}
\usepackage{cite}
\usepackage{makecell}

\usepackage{amsmath,amssymb,amsfonts}
\usepackage{algorithm}
\usepackage{algpseudocode}
\usepackage{graphicx}
\usepackage{textcomp}
\usepackage{subcaption}
\usepackage{booktabs}
\usepackage{diagbox}
\usepackage{tabularx}
\usepackage{setspace}
\usepackage{color}
\usepackage[colorlinks,linkcolor=blue]{hyperref}
\usepackage{cleveref}

\definecolor{newcolor}{rgb}{.8,.349,.1}

\journal{Medical Image Analysis}

\begin{document}

\verso{\textit{Chaoyu Chen et~al.}}

\begin{frontmatter}

\title{FetusMapV2: Enhanced Fetal Pose Estimation in 3D Ultrasound}%

\author[1,2]{Chaoyu Chen\fnref{fn1}}
\author[1,2]{Xin Yang\fnref{fn1}}
\author[1,2]{Yuhao Huang}
\author[3]{Wenlong Shi}
\author[3]{Yan Cao}
\author[1,2]{Mingyuan Luo}
\author[3]{Xindi Hu}
\author[5,6]{Lei Zhu}
\author[7]{Lequan Yu}
\author[8]{Kejuan Yue}
\author[1,2]{Yuanji Zhang}
\author[4]{Yi Xiong}
\author[1,2]{Dong Ni\corref{cor1}}
\ead{nidong@szu.edu.cn}
\author[9]{Weijun Huang\corref{cor1}}
\ead{hwjun1716@163.com}
\cortext[cor1]{Corresponding author. {$^{1}$}Authors contributed equally.}

\address[1]{National-Regional Key Technology Engineering Laboratory for Medical Ultrasound, School of Biomedical Engineering, Health Science Center, Shenzhen University, Shenzhen, China}
\address[2]{Medical Ultrasound Image Computing (MUSIC) Laboratory, Shenzhen University, Shenzhen, China}
\address[3]{Shenzhen RayShape Medical Technology Co., Ltd, Shenzhen, China}
\address[4]{Department of Ultrasound, Luohu People's Hosptial, Shenzhen, China}
\address[5]{The Hong Kong University of Science and Technology (Guangzhou), Nansha, Guangzhou, China}
\address[6]{The Hong Kong University of Science and Technology, Hong Kong SAR, China}
\address[7]{Department of Statistics and Actuarial Science, The
University of Hong Kong, Hong Kong SAR, China}
\address[8]{Hunan First Normal University, Changsha, China}
\address[9]{Department of Medical Ultrasonics, The First People's Hospital of Foshan, Foshan, China}

\received{*****}
\finalform{*****}
\accepted{*****}
\availableonline{*****}
\communicated{S. Sarkar}

\begin{abstract}
Fetal pose estimation in 3D ultrasound (US) involves identifying a set of associated fetal anatomical landmarks. Its primary objective is to provide comprehensive information about the fetus through landmark connections, thus benefiting various critical applications, such as biometric measurements, plane localization, and fetal movement monitoring. However, accurately estimating the 3D fetal pose in US volume has several challenges, including poor image quality, limited GPU memory for tackling high dimensional data, symmetrical or ambiguous anatomical structures, and considerable variations in fetal poses. In this study, we propose a novel 3D fetal pose estimation framework (called \textbf{FetusMapV2}) to overcome the above challenges. Our contribution is three-fold. First, we propose a heuristic scheme that explores the complementary network structure-unconstrained and activation-unreserved GPU memory management approaches, which can enlarge the input image resolution for better results under limited GPU memory. Second, we design a novel \emph{Pair Loss} to mitigate confusion caused by symmetrical and similar anatomical structures. It separates the hidden classification task from the landmark localization task and thus progressively eases model learning. Last, we propose a shape priors-based self-supervised learning by selecting the relatively stable landmarks to refine the pose online. Extensive experiments and diverse applications on a large-scale fetal US dataset including 1000 volumes with 22 landmarks per volume demonstrate that our method outperforms other strong competitors. 

\end{abstract}


\end{frontmatter}



\begin{figure*}[htb]
	\centering
	\includegraphics[width=0.95 \linewidth]{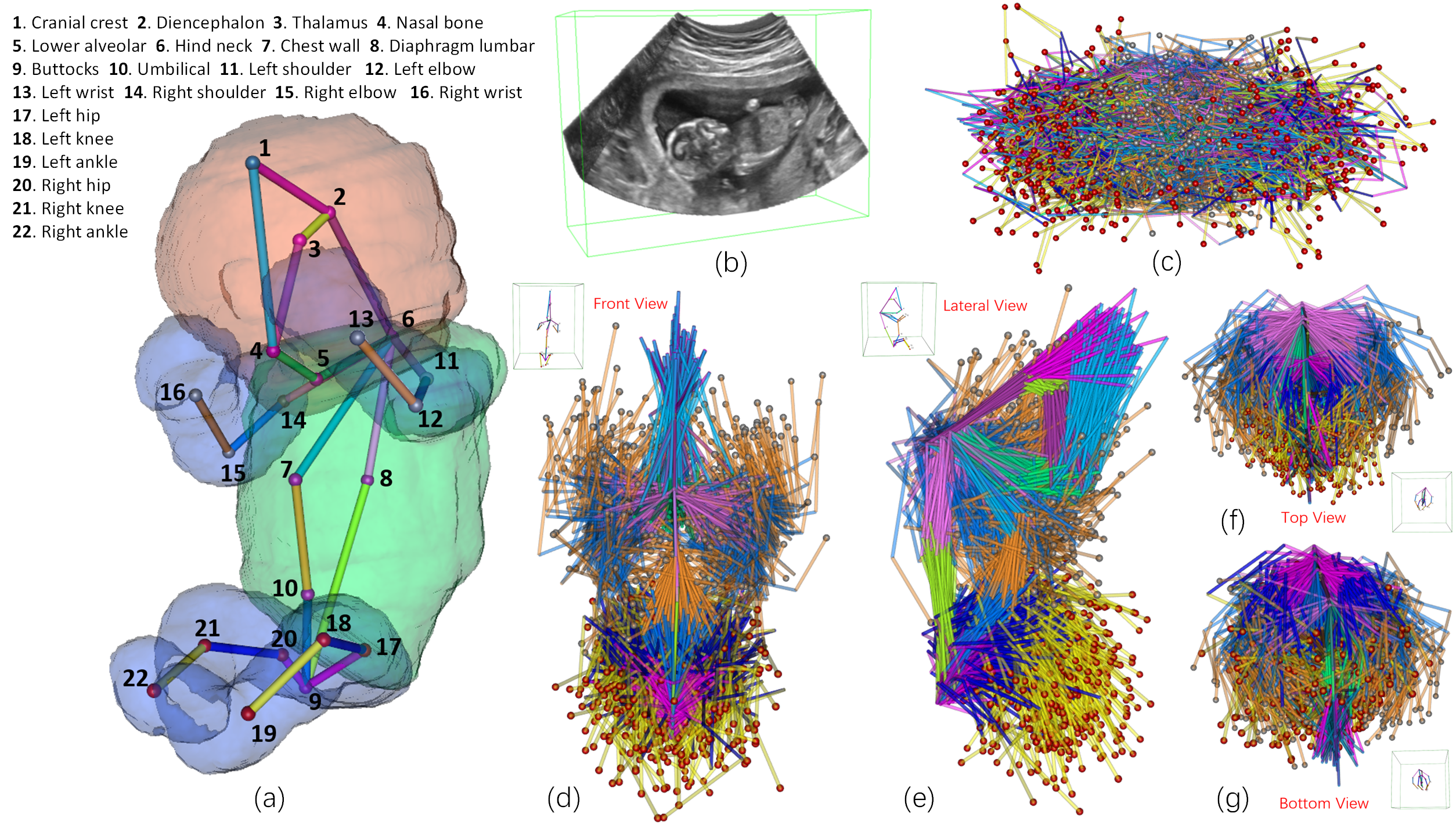}
	\caption{(a) An example of 3D fetal pose with 22 landmarks. (b) A representative 3D US data volume. (c) The sampled 500 fetal poses in our dataset. (d)-(g) Different views of 500 fetal poses after landmark-based registration. For (c)-(g), red landmarks correspond to the ankle, and gray landmarks correspond to the wrist. From (c)-(g), we discovered that the pose shift of the fetus is remarkably significant.}
	\label{fig:Instance}
	\vspace{-0.3cm}	
\end{figure*}\par

\section{Introduction}
Ultrasound (US) plays a crucial role in medical diagnosis, especially prenatal screening due to its advantages of non-radiation, low cost, and real-time imaging~\citep{reddy2008prenatal, noble2006ultrasound}. In the routine prenatal examination, using the 3D US probe to obtain the whole fetus in one shot can provide important spatial information which is not contained in the 2D US. 
Hence, sonographers can monitor a baby's global situation better.\par

Pose estimation originates from the computer vision field. 
These approaches are primarily based on deep learning (DL) framework~\citep{cv_pose_2d_2,cvpose_2d,cv_pose_3d}. 
The core purpose of pose estimation is to predict the spatial locations of a person's vital body joints (i.e., landmarks including elbow, knee, etc.) and connect the landmarks in a predefined method to develop the pose. Pose estimation has applications in many fields, such as human-computer interaction~\citep{human_computer_interaction}, action recognition~\citep{action_reg}, and motion capture~\citep{motion_cap}. \par

In response, we introduced and explored pose estimation in fetal analysis to assist physicians in performing various valuable clinical tasks, such as bio-metric measurement~\citep{alison_measure, fetal_3d_measurement, fetal_3d_measurement_2}, plane localization~\citep{fetal_plane_3d, yang2021searching}, and fetal movement monitoring~\citep{xu2019fetal}. 
Fig.~\ref{fig:Instance}(a) visualizes a typical example of the 3D fetal pose in our task. 
The fixed connections among 22 landmarks provide a comprehensive and systematic representation of the fetal pose in 3D space. 
Besides, via the 3D fetal landmarks, several important indicators of fetal growth and development (e.g., crown-lump length and femur length) can be calculated in one shot, which is more convenient than using 2D images. The reason is that bio-metric measurements in 2D are usually approximated on selected standard planes, and the acquisition of standard planes is experience-dependent and time-consuming. 
Thus, fetal pose estimation in 3D US has the potential to provide rich information for prenatal diagnosis. 
The goal of this study is to develop an automated system for detecting complex landmarks to further facilitate the fetal pose estimation task.\par

However, as shown in Fig.~\ref{fig:Instance}, there remain several challenges for pose estimation in the 3D US: 1) limited GPU resources constrain the resolution of the input in DL frameworks, especially analyzing poor-quality and high-dimensional US volume (see Fig.~\ref{fig:Instance}(b)). In addition, if the resolution of the input is simply reduced to a compromise with limited resources, the volume will produce more artifacts, resulting in inferior image quality and further impairing the model's performance. Thus, a primary concern of this study is to balance the resolution of the input fetal volume under limited GPU resources.
2) fetal limbs are anatomically similar, which can confuse network prediction. 
This issue is further complicated by the left-right crossover resulting from the movement of the limbs (Fig.~\ref{fig:Instance}(d-g)). 
Moreover, in the early development of the fetus, the accurate localization of landmarks of the head and trunk can be challenging because of the ambiguity of the surrounding tissues and anatomy.
3) the fetal poses are diverse and complex (see Fig.~\ref{fig:Instance}(c)). However, collecting sufficient training samples that cover all possible poses is unfeasible. This presents a significant challenge in training a model that, despite limited data, can generalize effectively to unseen poses.\par
Previous studies have focused on 2D landmark localization~\citep{sofka2017fully, payer2016regressing, newell2016stacked, cao2018openpose} and pose estimation~\citep{NI2023102654}, and these frameworks may be unsuitable for processing volume with high dimensions. Constrained by GPU resource limitations, some approaches depend on cropped patches~\citep{xu2019fetal, xu20203d}, which can result in the loss of global information of the whole fetus. Because of the complexity of the fetal pose, the methods that locate landmarks only in parts of the fetal body~\citep{chen2020region, huang2018omni} may experience performance degradation when extended to analyze the entire fetal pose. In this study, we propose a novel 3D fetal pose estimation framework (\textbf{FetusMapV2}) to analyze the whole fetus comprehensively. 
The main contributions are as follows:
\begin{itemize}
\item[1)]
We propose a heuristic GPU memory management scheme to enable the model to handle high-resolution inputs under limited resources. It is essential for obtaining high-quality images and achieving accurate pose estimations.

\item[2)]
We design a novel \emph{Pair Loss} that effectively separates the classification task from the localization task, thereby alleviating landmark confusion issues and further improving model's resistance to interference in estimating pose.

\item[3)]
We introduce a shape prior-based self-supervised learning (SSL) framework to guide the refinement of initial predictions during testing. This approach enhances the generalizability of the model in unseen poses.
\end{itemize}
\textbf{Differences from our conference work.} 
In our previous MICCAI-version study~\citep{yang2019fetusmap}, we proposed a fetal pose estimation framework named \textbf{FetusMap} to detect 16 landmarks in 3D US. Besides, we adopted the structure-unconstrained gradient check-pointing (GCP) strategy to save GPU memory, and SSL to provide a proxy for the online refinement of specific landmarks, i.e., the four limbs. 
In this study, \textbf{FetusMapV2} has several significant improvements and effective enhancements compared with \textbf{FetusMap}. (1) We add 6 additional important landmarks (16 to 22) of the head and trunk to characterize and model complex fetal pose accurately. (2) We substantially expand our dataset from 152 to 1000 cases to validate the effectiveness of our proposed framework. (3) We design a heuristic GPU memory management scheme that utilizes multiple complementary strategies. (4) We propose a novel \emph{Pair Loss} to alleviate the confusion issues effectively. (5) To accommodate various fetal poses, we redesign the SSL framework to refine volatile landmarks online.\par
Using the fetal pose obtained from detecting 22 key fetal landmarks, we can directly implement several clinical applications, and provide navigation for many advanced studies. For example, based on landmarks, some important measurement parameters can be calculated based on the length of two points (e.g., the crown-rump length). Various standard planes can be determined by multiple ($\geq 3$) points (e.g., mid-sagittal plane). Moreover, pose information can guide many downstream tasks, such as limb or trunk segmentation. Last, we believe that applying 3D pose estimation to 4D US images for fetal movement analysis might be intriguing.

\section{Related works}

\subsection{Deep learning in landmark detection and pose estimation}
Deep learning has been widely applied to medical image analysis tasks. Specifically, for US images, studies have focused on 2D images~\citep{golan2016fully,jang2017cnn,wang2020auto,huang2021flip,hu2021joint}, also, on 3D volumes~\citep{yang2017towards,zonoobi2018developmental,yang2021agent,yang2021searching}. Specifically, several studies have explored pose estimation through CNNs and one mainstream method is based on significant landmark detection.
This section summarizes related literature regarding 2D and 3D landmark detection in US volume and pose estimation.

\textbf{Landmark detection in 2D US images.} \citet{sofka2017fully} directly regressed the spatial coordinates of heart landmarks. 
Instead, some methods predict a Gaussian heatmap of each landmark and consider the position with the highest value as the final prediction~\citep{payer2016regressing, newell2016stacked, cao2018openpose}.
Although these methods are promising for 2D images, they may underperform when handling 3D volume directly. Because the 3D volume has a higher degree of freedom (DoF), increased variations in fetal poses make proper learning for the model challenging.\par

\textbf{Landmark detection in 3D US volumes.} Compared to 2D US image, 3D US volumes provide more effective spatial information, which is beneficial for pose estimation. \citet{huang2018omni} used a semi-supervised learning method to localize six fetal head landmarks. \citet{chen2020region} proposed an object detection-based method for detecting five fetal head landmarks. 
However, these methods handle only a part of the entrie fetus, that is, the  head.
Thus, they may fail to find more fetal landmarks with varying appearances.

\textbf{Pose estimation in fetus or infant.} \citet{xu2019fetal} used a 2-stage framework based on 3D UNet~\citep{cciccek20163d} to estimate the fetal pose by detecting 15 landmarks on 3D MRI. Recently,~\citet{xu20203d} proposed a conditional GAN based method to locate landmarks and further improve the robustness of estimating the fetal pose in 3D MRI. However, limited by GPU resources, patch-based methods were used to train the network, which lacked a vital global context. \citet{NI2023102654} proposed a semi-supervised framework for body parsing and pose estimation in 2D infant movement videos. Nevertheless, our 3D US volumes have lower quality and higher DoF than their infant videos, which may degrade the model's performance. 

\begin{figure*}[h]
\centering
\includegraphics[width=0.95\linewidth]{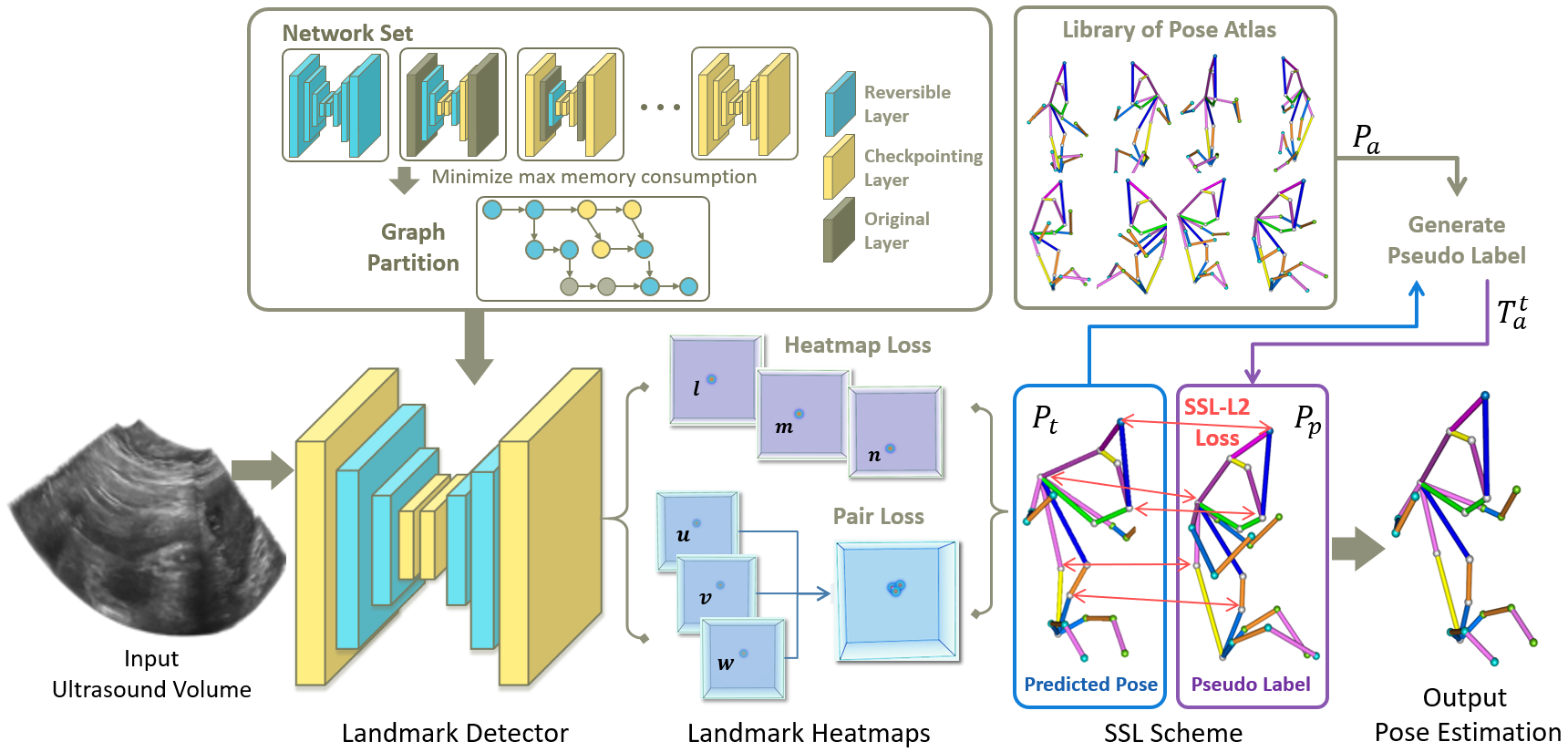}
\caption{The framework of our proposed FetusMapV2. Network Set indicates U-Net networks with different memory reduction module configurations. Graph Partition denotes the process of using our heuristic framework for network design. The Pose Library represents all the fetal poses in our training set. ${l, m, n, u, v, w}$ denote different landmarks. ${l, m, n}\in\{1, 2, 3, 4, 5, 6, 7, 8, 9, 10\}$, and $(u, v, w)\in\{(11, 12, 13), (14, 15, 16), (17, 18, 19), (20, 21, 22)\}$.}
\label{fig:framework}
\vspace{-0.3cm}	
\end{figure*}

\subsection{GPU memory resource management}
Training a network using full-size input guarantees the server the global context of volumes and further improves the model performance.
However, there is usually a trade-off between limited computing resources and resolution. Thus, several studies on methods to decrease memory usage while maintaining model precision have been conducted, which can ensure higher input resolution and better performance. There are currently two mainstream methods: dynamic sparse reparameterization techniques and recomputation algorithms. \par
Dynamic sparse reparameterization techniques~\citep{mostafa2019parameter} can make the model parameters sparse during training phase, which lowers the model parameter-related training memory.
Instead of using single-precision (32 bits) for all quantities, such as in standard training algorithms,
\citet{micikevicius2017mixed} advocated adopting half-precision (16 bits) to reduce memory requirements. 
However, when lower precision is used in some tasks, there is a risk of performance degradation.

\citet{chen2016training} first employed the gradient checkpoint (GCP) method for recomputation algorithms. To decrease the memory consumption of large networks during training, a structure-unconstrained GCP discards intermediate activations in the forward pass and recomputes them in the backward pass. Consequently, it can significantly decrease the memory requirements of networks without sacrificing their accuracy.
\citet{gomez2017reversible} proposed an activation-unreserved reversible layer (REV), an architecture designed to enable the backward pass without the need to retain intermediate activations during the forward pass. This result is achieved by carefully designing the layer's computations, such that they may be reversed precisely and efficiently. \citet{Etmann2020iUNets} designed IUNets to make upsample and downsample blocks invertible. 

Researchers have modified existing networks for more memory-efficient learning using these recomputation technologies.
\citet{gruslys2016memory} and \citet{pleiss2017memory} applied GCP to CNN blocks with the same pattern to decrease training memory consumption. 
To increase the resolution of the input volumes in the 3D segmentation task, \citet{brugger2019partially} designed a partially reversible 3D UNet, that replaced all Conv-BN-ReLU blocks with REV layers. 
However, they only modified their methods by inflexible finding and replacement, which may not utilize these recomputation technologies. \citet{blumberg2018deeper} combined GCP and REV to design a chain network, to obtain a deeper network for image quality transfer.
However, the design principle of their simple chain network structure cannot be applied to more complex networks, such as UNet. 
Recently, \citet{kusumoto2019graph} and \citet{kumar2019efficient} proposed frameworks for applying GCP to complex networks heuristically other than naive finding-and-replacing, however, their frameworks did not utilize REV layers.

In our study, we generalized the \emph{Kusumoto}'s algorithm~\citep{kusumoto2019graph}. It can heuristically combine multiple recomputation algorithms to manage the training memory for efficient usage.\par

\subsection{Strategy of exploiting landmark dependency}
Learning the relationship among different landmarks may have the potential to optimize the performance of both landmark detection and pose estimation tasks.
Some of the earliest methods leverage the shape priors to capture connections. Specifically, they adopted generative adversarial learning (GAN)~\citep{chen2017adversarial} and recurrent neural network (RNN)~\citep{xu2018less} based methods to model the dependency among landmarks and further improve the final results.
OpenPose proposed to using part-affinity fields to associate body parts with individuals, thus alleviating confusion between limbs (i.e., left and right)~\citep{cao2018openpose}.
\citet{xu2019fetal} used a Markov Random Field to correct landmark predictions.
One main disadvantage of their method is that they learn such landmark dependencies and update the model parameters during training.
Fetal poses, however, are arbitrary, diverse, and complex in clinical scenarios.
The training cases cannot cover all potential types of fetal poses. Thus, they may limit the model's generalizability. 
In this study, we proposed an SSL-based framework to optimize estimation results during testing phase.
We believe that online learning strategies can improve the robustness of these models.

\begin{figure*}[!ht]
\centering
\includegraphics[width=1.0\linewidth]{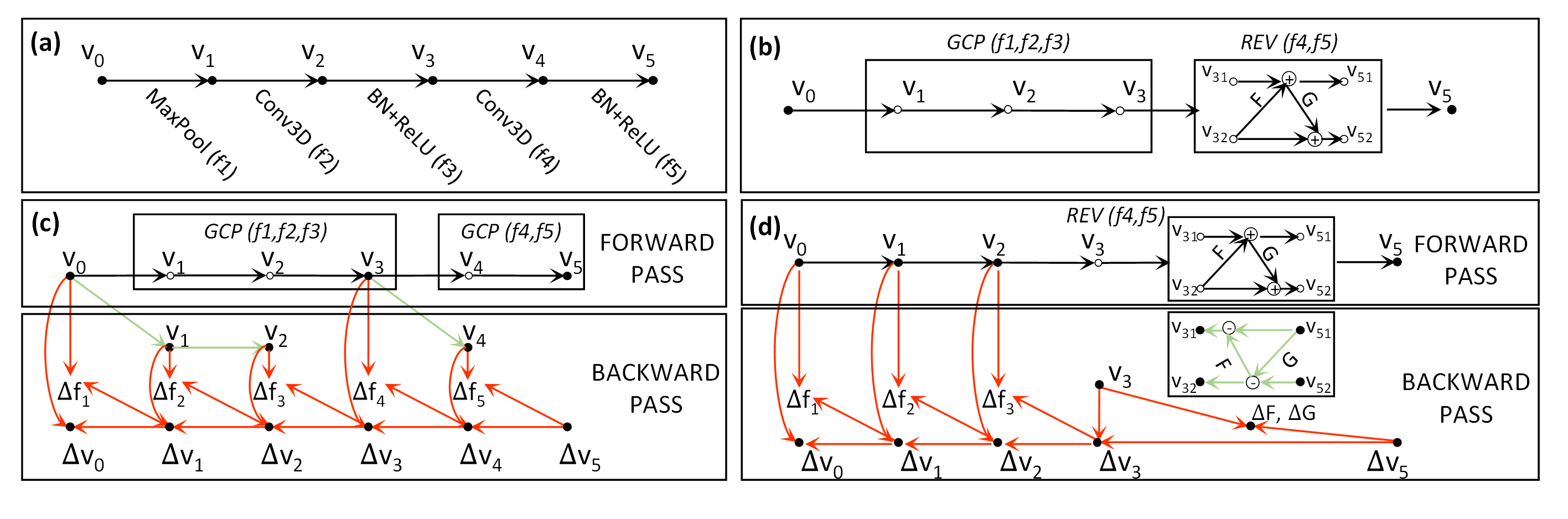}
\caption{{The principle of GCP and REV. (a) is the base block. (b) stands for the strategy on how to combine GCP and REV simultaneously. (c) and (d) show how GCP and REV are used dependently along with their computation in forward and backward passes. $V_0$-$V_5$ represent the feature vectors obtained from the last operations. Solid and hollow points represent retained and discarded feature vectors, respectively. $\Delta \cdot$ means the gradient of this element. Red arrows mean the gradient calculation, while green arrows indicate recalculating the discarded features during backpropagation.}}
\label{fig:cprev}
\vspace{-0.3cm}	
\end{figure*}

\section{Methodology}
\label{sec:method}

Fig.~\ref{fig:framework} illustrates the schematic representation of our method. 
The framework takes a fetal US volume as input and produces the 3D coordinates of anatomical key landmarks.
We exploit a memory-efficient landmark detector to enable high-resolution inputs for accurate prediction of each landmark. 
During training, we optimize the landmarks of different types by utilizing \emph{Pair Loss} to achieve more stable outputs. During testing, we equip SSL to enhance landmark prediction performance.

\subsection{GPU memory management for high-resolution input}
The capacity of a 3D convolutional neural network to process volumetric data is restricted by the memory hardware due to the considerable storage demand imposed by the intermediate activations and gradients during the backpropagation process.
Previous approaches usually divided the input volume into small patches or down-sampled it to reduce memory consumption. 
However, patch-based approaches have been observed to exhibit a deficiency in accounting for global context, while down-sampling-based techniques tend to generate a great number of artifacts. 
Memory-efficient training strategy has the potential to maximize input size, resulting in an overall decrease in landmark detection errors.
For example, model sparsity~\citep{mostafa2019parameter} and mixed-precision training~\citep{micikevicius2017mixed} have been validated to reduce parameter-related memory.
Nevertheless, they may cause performance degradation.
An alternative called recomputation can reduce memory with no impact on accuracy, which is suitable for our high-precision pose estimation task.

The recomputation algorithms are training memory management technologies that selectively discard intermediate activations during the forward pass and on-demand recompute them during the backward pass. Two common methods include structure-unconstrained GCP~\citep{chen2016training} and activation-unreserved REV~\citep{gomez2017reversible}.
Fig.~\ref{fig:cprev}(a) shows a 3D UNet encoder block, which can be memory-efficient if equipped with GCP and REV.
The structure-unconstrained GCP discards the block's intermediate activations in forward propagation and recovers them from the reserved input in backward propagation. Specifically, in Fig.~\ref{fig:cprev}(c), GCP is applied to $v_1\rightarrow v_2\rightarrow v_3$ and $v_4\rightarrow v_5$ passes. $v_1$, $v_2$, and $v_4$ are discarded in forward pass and will be recovered by $v_0$ and $v_3$ in backward pass.
The activation-unreserved REV takes a more aggressive strategy without storing either input or intermediate activation since they can be calculated from the output in backward pass. But the REV block needs input and output in the same form. Its forward and backward passes can be respectively expressed as:
\begin{equation}
\label{Eq:rev1}
\begin{split}
\begin{cases}
v_{31}, v_{32} = \text{split}(v_3) \\
v_{51} = v_{31} + F(v_{32}) \\
v_{52} = v_{32} + G(v_{51}) \\
v_5 = \text{concat}(v_{51}, v_{52})
\end{cases}
\end{split}
\quad
\begin{split}
\begin{cases}
v_{51}, v_{52} = \text{split}(v_5) \\
v_{32} = v_{52} - G(v_{51}) \\
v_{31} = v_{51} - F(v_{32}) \\
v_3 = \text{concat}(v_{31}, v_{32})
\end{cases}
\end{split}
\end{equation}
where {the \textit{split} operation represents to divide a feature evenly into two features in the channel dimension.} 
$F$ and $G$ can be any arbitrary function. $v_{31}$, $v_{32}$, $v_{51}$ and $v_{52}$ partitioned from input $v_3$ and output $v_5$ are all of equal shape. Input $v_3$ and intermediate activations can all be discarded in forward computation.
Specifically, as shown in Fig.~\ref{fig:cprev}(d), $v_3\rightarrow v_4\rightarrow v_5$ pass is replaced by REV, which takes input $v_3$ and output $v_5$ of the same shape.
As described above, GCP and REV both skillfully reduce training activation memory. They each have benefits and drawbacks. GCP can be applied to any network block but the input cannot be discarded. REV can discard its input, but it can only replace the special block with the same form of input and output. Fig.~\ref{fig:cprev}(b) illustrates a strategy to combine GCP and REV, which is more efficient than using them alone. GCP and REV are applied to $v_1\rightarrow v_2\rightarrow v_3$ and $v_3\rightarrow v_4\rightarrow v_5$ passes, respectively. In forward pass, $v_1$ and $v_2$ are discarded in GCP block, while $v_3$ and intermediate activations are discarded in REV block.

\begin{figure*}[tbp]
	\centering
	\includegraphics[width=1.0\linewidth]{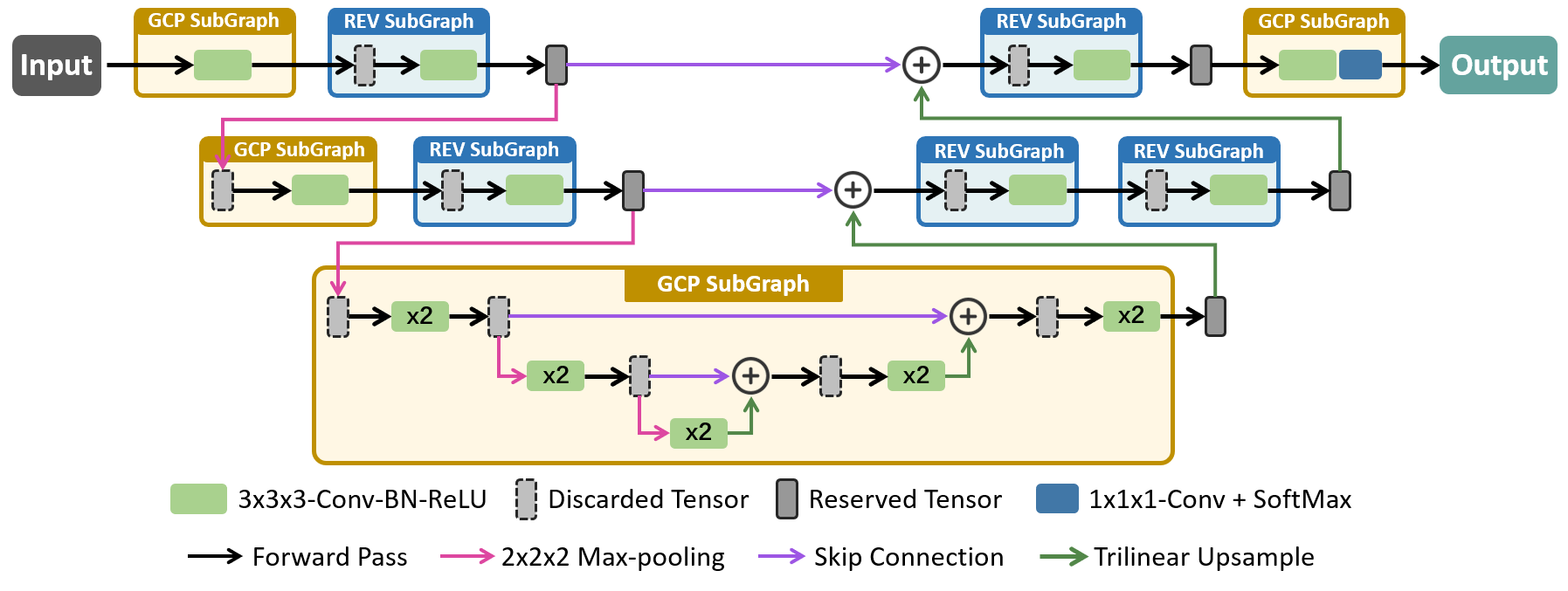}
	\caption{3D UNet equipped with GCP and REV. In the GCP/REV subgraph, any intermediate tensors generated by forward propagation are discarded to save GPU resources. This approach reserves only a small fraction of the necessary tensor during forward propagation.}
	\label{fig:cprev_arch}
	\vspace{-0.3cm}	
\end{figure*}

\begin{figure}[thb]
\centering
\includegraphics[width=1.0\linewidth]{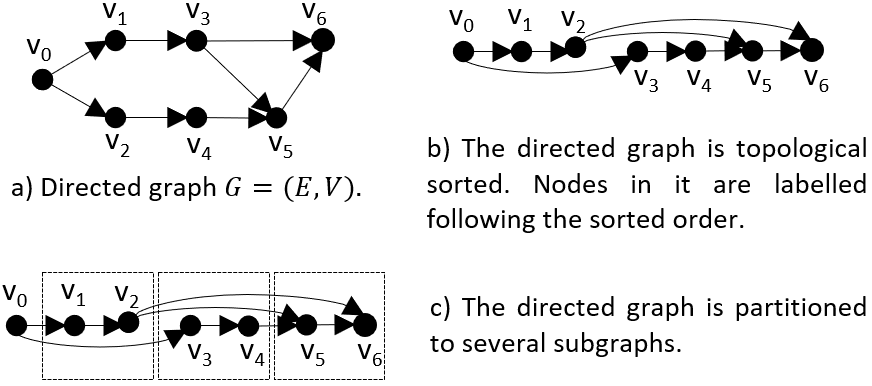}
\caption{A directed graph is sorted, labeled, and partitioned.}
\label{fig:graph_partition}
\vspace{-0.3cm}	
\end{figure}

In this study, we propose a heuristic memory management framework that explores the complementary GCP and REV to optimize memory usage and training efficiency.
Specifically, the architecture of a neural network can be represented as a directed graph $G=(V,E)$ (for example in Fig.~\ref{fig:graph_partition}(a)). The node $v\in V$ are activation, input, and output node in the network. There is an edge $(v_i, v_j)\in E$ if $v_i$ is an input of direct computation for $v_j$.
If applying topological sorting to $G$, $G$ will be flattened to nodes in order. Following the order, nodes in $G$ can be labeled in sequence as shown in Fig.~\ref{fig:graph_partition}(b). 
For any subgraph $S\subset G$, we call the node with the largest ordering in $S$ the output node $v_{o}(S)$.
A subgraph $S\subset G$ is valid if no edge connects a node in $S-\{v_o(S)\}$ to a node in $G-S$ (for example in Fig.~\ref{fig:graph_partition}(c)).
For a valid subgraph partition $G_S=\{S_i|\bigcup^{M}_{i=1}S_i=G\}$, we define the subgraph execution function $f:G_S\rightarrow\{1,2,3\}$ to determine each subgraph's recomputation algorithm, where $1$, $2$, and $3$ mean no operation, applying GCP, and replacing with a REV, respectively.
We define the max memory consumption as $M(G_S,f)$.
We formalize our task for finding the optimal memory management strategy as follows: given a neural network with graph representation $G$, find the valid graph partition $G_S$ and subgraph execution function $f$ that minimizes $M(G_S,f)$.
Our heuristic memory management framework to solve the formalized problem can be divided into three steps as follows:
\begin{enumerate}
\item Use \emph{Kusumoto}'s algorithm~\citep{kusumoto2019graph} to find a valid partition strategy for graph $G$ that minimizes memory consumption during training  when only using GCP. The result is a valid partition strategy $G_S$ and a subgraph execution function $f:G_S\rightarrow\{1,2\}$.
\item Traverse graph $G$ to find a subgraph partition $R_S$ in which each subgraph could be replaced by a REV block. The result is $R_S=\{R_i|\bigcup^{m}_{i=1}R_i=G\}$, where $R_i$ is represented by a \emph{Conv}-\emph{BN}-\emph{ReLU} pass.
\item Traverse $R_S$ and determine if each $R_i$ should be replaced by a REV block. For $G_S=\{S_i|\bigcup^{M}_{i=1}S_i=G\}$ and $f:G_S\rightarrow\{1,2\}$, if $R_i\in R_S$ is replaced by a REV block, $G_S$ could be converted to $G_S^{'}=\{R_i,S_0^{'},S_1^{'},\cdots,S_M^{'}\}$, where $S_i^{'}=\{v|v\in S_i,v\notin R_i\}$, and $f$ could be converted to $f^{'}:G_S^{'}\rightarrow\{1,2,3\}$, where $f^{'}(R_i)=3$ and $f^{'}(S_i^{'})=f(S_i)$. If all converted subgraphs $S_i^{'}$ are valid and conversion saves memory ($M(G_S^{'},f^{'})<M(G_S,f)$), $R_i$ should be replaced with a REV block.
\end{enumerate}

We build a fetal pose landmark detector by applying our heuristic memory management framework to the well-known 3D UNet~\citep{cciccek20163d}, as shown in Fig.~\ref{fig:cprev_arch}. 

\subsection{Pair Loss for learning confusion landmarks}

The key landmarks with symmetrical and similar anatomical structures can confuse the learning of deep models, especially in fetal limb estimation.
For example, the left-right relationship between the limbs, is likely to be confused in prediction~\citep{guo2019triple, choi2018subcategory}.
We consider that this is because the common landmark localization task couples a hidden classification task, which complicates the whole learning process.
Therefore, we designed a \emph{Pair Loss} to separate the localization task from the classification task. Specifically, it classifies key landmarks using prior knowledge, which can reduce the probability of confusion. 

{In this work, we use the Gaussian distribution to convert the annotated 3D coordinates ($g_i$) into the 3D heatmap (as shown in Fig.~\ref{fig:framework}) of ground-truth ($H(g_i)$), following as :
        \begin{equation}
            H(x,y,z) = \frac{1}{(\sqrt{2\pi}\sigma)^3 } e^{-\frac{[(i-x)^2+(j-y)^2+(k-z)^2]}{2\sigma^2}}
        \end{equation}
where $(x, y, z)$ is the annotated 3D coordinates, $(i, j, k)$ is the 3D coordinates of a point in the heatmap, and $\sigma = 2$ in this study. We convert the predicted heatmap ($H(p_i)$) to 3D point coordinates($p_i$) using a simple $Argmax$ formula.
}

\begin{figure}[htb]
\centering
\includegraphics[width=0.8 \linewidth]{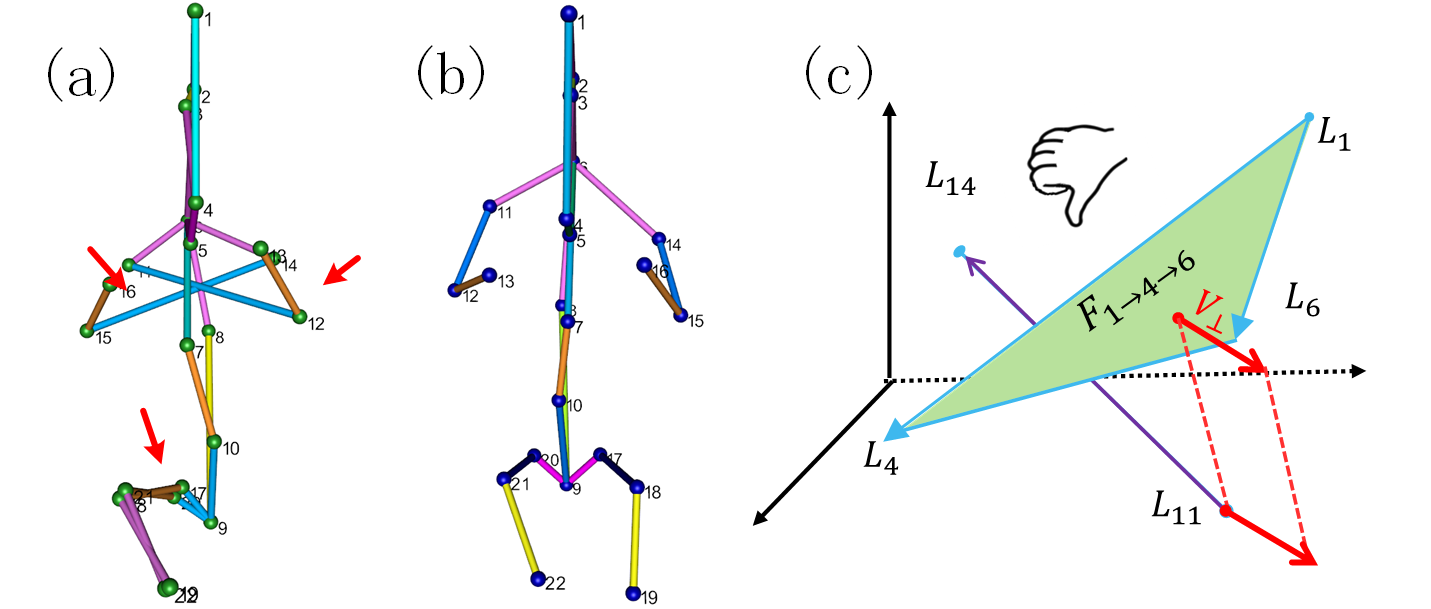}
\caption{(a) Prediction without \emph{Pair Loss}. (b) Ground truth. (c) Distinguish left-right relationships using spatial geometry knowledge.}
\label{fig:grid_regis}	
\end{figure}

As shown in Fig.~\ref{fig:Instance}(a), all landmarks can be divided into five groups: head and trunk (\emph{ht}, $1,2,\cdots,10$), left arm (\emph{la}, $11,12,13$), right arm (\emph{ra}, $14,15,16$), left leg (\emph{ll}, $17,18,19$), and right leg (\emph{rl}, $20,21,22$). {The estimated and ground-truth heatmap for left arm are defined as $P_{\textit{la}}=\left\{H(p_i)|i=11,12,13\right\}$ and $G_{\textit{la}}=\left\{H(g_i)|i=11,12,13\right\}$.} And so on for the other four groups.
We divide \emph{Pair Loss} ($L_{pair}$) into two parts:
\begin{equation}
L_{pair}(P_{\textit{l}},P_{\textit{r}},G_{\textit{l}},G_{\textit{r}})=L_1+L_2 \ ,
\label{Equ.hc}
\end{equation}
where \emph{l}/\emph{r} represents left/right side landmarks, respectively.  {It is worth noting that, the \emph{Pair Loss} focuses on optimizing the left and right limbs of the fetal pose to alleviate the confusion problem.} The first part of \emph{Pair Loss} is defined as:
\begin{equation}
L_1=\min(K(P_{\textit{l}},G_{\textit{l}})+K(P_{\textit{r}},G_{\textit{r}}),K(P_{\textit{l}},G_{\textit{r}})+K(P_{\textit{r}},G_{\textit{l}})) \ ,
\label{Equ.hc1}
\end{equation}
where $K(P,G)$ represents \emph{Kullback-Leibler} divergency to measure the distance between predicted and ground-truth probability distribution in heatmap:
\begin{equation}
\begin{split}
K(P,G)=-\sum_{H(p_i)\in P,H(g_i)\in G}\log\left(\frac{e^{H(p_i)}}{\sum_{j\in P}{e^{j}}}\right)\odot H(g_i),
\end{split}
\label{Equ.HM}
\end{equation}
in which $\odot$ denotes element-wise product. 
In $L_1$, we separate the task of distinguishing whether the group of landmarks belongs to the left or right side from the task of localization. {Therefore, we detect two groups of landmarks without having to confirm which represents the left or right side, that reduces the difficulty of optimization.} For instance, the landmarks of elbows and wrists shown in Fig.~\ref{fig:grid_regis}(a) are mislocalization. The traditional loss forces the network to optimize the mislocalization of elbows and wrists, while $L_1$ will only optimize the landmarks of the shoulders.
The second part of $L_{pair}$ is defined as:
\begin{equation}
L_2=-\sum_{
\mbox{\tiny$
\begin{array}{c}
H(p^\textit{l}_i)\in P_{\textit{l}},H(p^\textit{r}_i)\in P_{\textit{r}},\\
H(g^\textit{l}_i)\in G_{\textit{l}},H(g^\textit{r}_i)\in G_{\textit{r}}
\end{array}$}
}\log\left(\frac{e^{H(p^\textit{l}_i)}}{\sum_{j\in P_{\textit{l}}}{e^{j}}}+\frac{e^{H(p^\textit{r}_i)}}{\sum_{j\in P_{\textit{r}}}{e^{j}}}\right)\odot\left(H(g^\textit{l}_i)+H(g^\textit{r}_i)\right).
\label{Equ.hc2}
\end{equation}
$L_2$ allows the predicted joint distribution of landmarks from two sides closer to the ground truth, which makes the predicted two landmarks will not appear on the same side such as the legs shown in Fig.~\ref{fig:grid_regis}(a). With these two parts, \emph{Pair Loss} ensures the plausibility of the predicted fetal pose, which makes the prediction closer to the ground truth (Fig.~\ref{fig:grid_regis}(b)).

{Then, for the predicted four groups (\emph{la}, \emph{ra}, \emph{ll} and \emph{rl}) of landmarks, the detector only ensures that each group of landmarks belongs to the same side, without distinguishing the left and right relationships between these groups of landmarks.} Thus, we should employ prior knowledge about spatial geometry to determine the left-right relationship of these groups of landmarks. {What we want to emphasize is that since the group of landmarks belong to the same side, we only need to determine the spatial position of any landmark in the group.}
Specifically, for shoulders (left, $11$, right, $14$) as shown in Fig.~\ref{fig:grid_regis}(c), we used non-collinear landmarks corresponding to the cranial crest ($1$), nasal bone ($4$), and hind neck ($6$) to determine the left-right relationship (Fig.~\ref{fig:grid_regis}(c)):
\begin{equation}
Ang = V_{14} \times V_{16} \cdot V_{s_1s_2},
\left\{
\begin{aligned}
s_1=11,s_2=14,&\text{  if } Ang \leq 0 \\
s_1=14,s_2=11,&\text{  otherwise},
\end{aligned}
\right.
\label{Equ.right_left}
\end{equation}
where $\times$ and $\cdot$ indicate cross product and dot product, respectively. {In the inference phase, our landmark detector only outputs two shoulder landmarks ($s_1$, $s_2$) without confirming the left-right relationship.
Using Eq.~\ref{Equ.right_left}, we can confirm the left-right relationship of $s_1$ and $s_2$ based on $Ang$.} Afterward, the left-right relationship of arms can be determined. The same method can be applied to the legs.

For left arm (\emph{la}), right arm (\emph{ra}), left leg (\emph{ll}), and right leg (\emph{rl}) landmarks, we use $L_{pair}$ in Eq.~\ref{Equ.hc} as loss function. For head and trunk (\emph{ht}) landmarks, we use $K$ in Eq.~\ref{Equ.HM} as the loss function. Therefore, the overall loss function can be formalized as:
\begin{equation}
\begin{split}
L=L_{pair}(P_{la},P_{ra},G_{la},G_{ra})+L_{pair}(P_{ll},P_{rl},G_{ll},G_{rl})+K(P_{ht},G_{ht}).
\label{Equ.loss_all}
\end{split}
\end{equation}

The overall loss function can effectively alleviate the landmark confusion problem, thus greatly improving the localization performance of the network, as elaborated in Section~\ref{sec:experiment}.
\subsection{Self-supervised learning (SSL) for online refinement}
{In the task of fetal pose estimation, due to the quality of US volumes, it may make several structures similar to the surrounding tissues (Fig.~\ref{fig:Instance}(a)-(b)). Thus, it is difficult for deep models to identify them during prediction. Not only that, the diversity and uncontrollability of fetal poses (Fig.~\ref{fig:Instance}(c)) make it impossible to collect all pose cases to train the model. It is a generalization problem that needs to be solved. Thankfully, based on the shape prior of clinical anatomy, we discover that the relative positional relationship between head and trunk landmarks is relatively stable, while the landmarks of the limbs have a significant range of motion (Fig.~\ref{fig:Instance}(d-f)). Thus, ambiguous landmarks of head and trunk can be reasonably inferred based on other stable landmarks (For example, even if the fetus rotates its head, the positions among the landmarks of the head are relatively stable).

Therefore, we designed a shape prior based \emph{SSL} framework to refine the landmarks of head and trunk during testing.} It can not only improve the robustness of ambiguous landmark prediction results, but also enhance the generalization of the model to unseen poses. In particular, we use the training poses as our shape prior library $S$.\par
The steps for online refinement are as follows:
\begin{enumerate}
\item Register $S$ to the predicted pose $P_t$;
\item Select several poses $P_a$ from the registered $S$;
\item Generate pseudo-label $P_p$ to online learning.
\end{enumerate}

\begin{table*}[tb!]\tiny
\setlength{\abovecaptionskip}{-0.1cm}
\caption{{Detailed descriptions of the different methods involved in the experiments (P and V denote the input mode of Patch and Whole Volume, respectively).}} 
\label{table_METHOD_detail} %
\begin{center}
\renewcommand\arraystretch{1.0}
\resizebox{\textwidth}{!}{
\begin{tabular}{c|c|c|c|c|c|c}
\Xhline{0.8pt}
\multirow{3}{*}{ \centering \textbf{{Methods}}} & \multirow{3}{*}{\textbf{{Network}}} &\multicolumn{2}{c|}{\textbf{{GPU}}} & \multirow{3}{1.5cm}{\centering \textbf{{Maximum}} \\ \textbf{{Input Size (Mode)}}} &\multirow{3}{*}{\textbf{{Pair Loss}}} & \multirow{3}{*}{\textbf{{SSL}}}\\
\cline{3-4} 
& & \multirow{2}{1cm}{\centering \textbf{{Memory}} \\ \textbf{{Limit}}}& \multirow{2}{1cm}{\centering \textbf{{Management}} \\ \textbf{{Scheme}}}& & &\\
& & & & & &\\
\hline

{UPatch}   &{UNet}& {8 GB}&{×}& {[176, 112, 112] (P)}&{×}& {×}\\
{UNet}  &{UNet}& {8 GB} &{×} &{[176, 112, 112] (V)} &{×}&{×}\\
{URNN} & {UNet+RNN}& {8 GB} &{×} &{[176, 112, 112] (V)} & {×} & {×}\\ 
{UGAN} & {UNet+GAN}&  {8 GB}& {×} &{[176, 112, 112] (V)} & {×} & {×} \\
{UPSSL} &{UNet} & {8 GB}& {×} &{[176, 112, 112] (V)} & {\checkmark} &{\checkmark} \\
{UGCP} & {UNet} &{8 GB}  & {GCP}  &{[192, 112, 128] (V) }& {×} &{×}\\
{UREV}  & {UNet} & {8 GB} &{REV}  &{[192, 128, 128] (V)}& {×} & {×} \\
{UGCPREV}  & {UNet}& {8 GB} & {GCP+REV}& {[208, 128, 144] (V)} &{×} &{×}  \\
{FetusMap}  & {UNet} &{8 GB} & {GCP} & {[192, 112, 128] (V)}& {×} & {\checkmark} \\
{FetusMapV2} & {UNet} & {8 GB} &{GCP+REV}&{[208, 128, 144] (V)} & {\checkmark} & {\checkmark} \\
\Xhline{0.8pt}

\end{tabular}}
\end{center}
\end{table*}

\begin{table*}[tb!]
\setlength{\abovecaptionskip}{-0.1cm}
\caption{Results of Euclidean Distance error for different methods.}
	\begin{center}	
		\renewcommand\arraystretch{1.6}
		\resizebox{\textwidth}{!}{
			\begin{tabular}{c|c|c|c|c|c|c|c|c|c|c|c|c|c|c|c|c|c|c|c|c|c|c|c}
                \Xhline{1.5pt}
				\multirow{2}{*}{\textbf{Methods}}  &\multicolumn{23}{c}{\Large\textbf{ED error($mm$) $\downarrow$ }} \\
				\cline{2-24} 
				& {\textbf{$L_{1}$}} & {\textbf{$L_{2}$}} & {\textbf{$L_{3}$}} & {\textbf{$L_{4}$}} 
				& {\textbf{$L_{5}$}} & {\textbf{$L_{6}$}} & {\textbf{$L_{7}$}} & {\textbf{$L_{8}$}} 
				& {\textbf{$L_{9}$}} & {\textbf{$L_{10}$}} & {\textbf{$L_{11}$}} & {\textbf{$L_{12}$}} 
				& {\textbf{$L_{13}$}} & {\textbf{$L_{14}$}} & {\textbf{$L_{15}$}} & {\textbf{$L_{16}$}} 
				& {\textbf{$L_{17}$}} & {\textbf{$L_{18}$}} & {\textbf{$L_{19}$}} & {\textbf{$L_{20}$}} 
				& {\textbf{$L_{21}$}} & {\textbf{$L_{22}$}} & {\textbf{$mean$}}\\
				\hline
                    \textbf{{UPatch}} & {3.89} & {2.44} & {2.02} & {2.56} & {2.03} &{4.52}  & {4.10} & {4.26} & {4.44} & {4.42} & {5.82} & {6.79} & {7.45} & {4.70} & {8.73} & {7.87} & {4.82} & {5.40} & {8.42} & {4.24} & {4.73} & {7.92} & {5.07} \\

				\textbf{UNet} & 4.07 & 2.46 & 2.30 & 3.14 & 
				2.16 & 3.97 & 4.04 & \textbf{3.51} & 4.93 & 4.22 &     4.59 & 5.03 & 5.59 & 4.04 & 4.80 & 5.62 & 3.59 &       4.27 & \textbf{5.39} & 3.71 & 4.14 & 5.55 & 4.14 \\
				
				\textbf{URNN} & 4.62 & 3.24 & 2.60 & 3.25 & 
				2.32 & 4.48 & 4.11 & 3.89 & 4.53 & 4.43 & 4.90 & 
				7.26 & 6.88 & 4.33 & 6.59 & 6.89 & 4.35 & 5.14 & 
				5.64 & 3.90 & 4.72 & 5.67 & 4.72   \\		
				
				\textbf{UGAN} & 4.11 & 2.40 & 1.99 & 2.79 & 
				2.13 & 3.99 & 4.13 & 3.95 & 4.18 & 4.62 & 3.67 & 
				3.97 & 5.90 & 3.00 & 3.98 & 5.58 & 3.43 & 4.31 & 
				5.64 & 3.41 & 3.93 & 6.04 & 3.96   \\	
				
				\textbf{UPSSL} & 3.56 & 2.51 & 2.07 & 
				2.59 & 2.14 & 3.85 & 4.05 & 3.75 & 4.29 & 4.18 & 
				3.81 & 4.16 & 4.91 & 3.45 & \textbf{3.55} & 5.22 & 3.09 & 
				3.82 & 5.51 & 3.14 & 3.97 & 5.47 & 3.78   \\
				
				\textbf{UGCP} & 4.07 & 2.56 & 2.10 & 2.47 & 
				2.04 & 3.83 & 3.88 & 3.80 & 4.47 & 4.53 & 3.65 & 
				4.50 & 5.06 & 3.13 & 4.21 & 5.78 & 3.16 & 4.22 & 
				5.77 & 3.02 & 3.96 & 5.55 & 3.90   \\	
				
				\textbf{UREV} & 3.62 & 2.38 & 1.98 & 2.56 & 
				2.06 & 3.84 & 4.29 & 3.73 & 4.16 & 4.50 & \textbf{3.17} & 
				4.57 & 5.50 & 2.95 & 4.29 & 5.21 & 3.44 & 3.95 & 
				5.76 & 3.19 & 3.64 & 5.79 & 3.84  \\	
				
				\textbf{UGCPREV} & 3.85 & 2.26 & 2.01 & 2.56 & 
				\textbf{2.02} & 3.90 & \textbf{3.68} & 3.67 & \textbf{3.73} & 4.14 & 3.62 & 
				4.20 & \textbf{4.87} & 3.13 & 3.80 & 5.14 & 3.29 & 3.90 & 
				5.50 & 3.20 & 4.01 & 5.66 & 3.73  \\
				
				\textbf{FetusMap} & 3.48 & 2.56 & 2.10 & 2.41 & 
				2.04 & 3.78 & 3.88 & 3.82 & 4.47 & 3.97 & 3.64 & 
				5.03 & 6.27 & 3.38 & 4.56 & 6.37 & 3.32 & 4.56 & 
				6.47 & 3.19 & 4.36 & 6.18 & 4.08  \\
				
				\textbf{FetusMapV2} & \textbf{3.32} & \textbf{2.21} & \textbf{1.91} & 
				\textbf{2.37} & 2.09 & \textbf{3.35} & 3.79 & 3.60 & 3.85 & \textbf{3.75} & 
				3.20 & \textbf{3.88} & 5.09 & \textbf{2.91} & 3.77 & \textbf{4.59} & \textbf{2.81} & 
				\textbf{3.41} & 5.46 & \textbf{2.66} & \textbf{3.05} & \textbf{5.13} & \textbf{3.46}  \\
				
                \Xhline{1.5pt}
		\end{tabular}}
		\label{tab:all_methods_dis}
	\end{center}
\end{table*}

\begin{table*}[tb!]
\setlength{\abovecaptionskip}{-0.1cm}
\caption{Results of AUC for different methods.}
	\begin{center}	
		\renewcommand\arraystretch{1.8}
		\resizebox{\textwidth}{!}{
			\begin{tabular}{c|c|c|c|c|c|c|c|c|c|c|c|c|c|c|c|c|c|c|c|c|c|c|c}
				\toprule[1.5pt] 
				\multirow{2}{*}{\textbf{Methods}}  &\multicolumn{23}{c}{\Large\textbf{AUC ($\%$)  $\uparrow$ }} \\
				\cline{2-24} 
				& {\textbf{$L_{1}$}} & {\textbf{$L_{2}$}} & {\textbf{$L_{3}$}} & {\textbf{$L_{4}$}} 
				& {\textbf{$L_{5}$}} & {\textbf{$L_{6}$}} & {\textbf{$L_{7}$}} & {\textbf{$L_{8}$}} 
				& {\textbf{$L_{9}$}} & {\textbf{$L_{10}$}} & {\textbf{$L_{11}$}} & {\textbf{$L_{12}$}} 
				& {\textbf{$L_{13}$}} & {\textbf{$L_{14}$}} & {\textbf{$L_{15}$}} & {\textbf{$L_{16}$}} 
				& {\textbf{$L_{17}$}} & {\textbf{$L_{18}$}} & {\textbf{$L_{19}$}} & {\textbf{$L_{20}$}} 
				& {\textbf{$L_{21}$}} & {\textbf{$L_{22}$}} & {\textbf{$mean$}}\\
				\hline
                     \textbf{{UPatch}} & {82.72} & {88.46} & {89.86} & {87.23} & {89.89} &{77.58}  & {79.39} & {78.95} & {79.85} & {78.42} & {73.37} & {72.34} & {67.56} & {78.12} & {66.92} & {67.98} & {78.86} & {74.04} & {65.34} & {79.82} & {76.68} & {68.40} &{77.35} \\
                    
				\textbf{UNet} & 80.08 & 87.69 & 88.77 & 84.46 &
				89.20 & 80.59 & 79.78 & 82.45 & 79.02 & 79.39 & 
				78.23 & 77.46 & 74.53 & 79.91 & 78.46 & 74.19 & 
				83.24 & 79.24 & 75.59 & 82.66 & 80.31 & 74.42 &
				80.44  \\
				
				\textbf{URNN} & 79.40 & 85.39 & 87.92 & 84.04 & 
				88.46 & 78.46 & 79.44 & 80.89 & 79.29 & 78.01 & 76.27 & 
				69.22 & 68.13 & 78.46 & 70.84 & 68.35 & 79.92 & 75.31 & 
				73.69 & 81.04 & 77.24 & 73.83 & 77.89  \\	
				
				\textbf{UGAN} & 80.57 & 88.00 & 90.04 & 86.15 & 
				89.43 & 80.20 & 79.53 & 80.46 & 79.65 & 77.55 & 82.79 & 
				81.92 & 73.64 & 85.36 & 81.47 & 75.03 & 83.88 & 80.27 & 
				75.32 & 84.38 & 81.34 & 72.85 & 81.36  \\	
				
				\textbf{UPSSL} & 83.00 & 88.29 & 89.67 &
				87.53 & 89.19 & 82.53 & 79.06 & \textbf{83.24} & 79.92 &
				78.98 & \textbf{84.79} & \textbf{84.19} & \textbf{79.05} & 82.93 & \textbf{84.35} &
				76.78 & 84.77 & 81.36 & 74.23 & 84.77 & 81.71 &
				72.86 & 82.42  \\
				
				\textbf{UGCP} & 81.30 & 87.20 & 
				89.55 & 87.64 & 89.74 & 81.00 & 80.59 & 81.39 & 78.60 &
				77.34 & 82.72 & 80.20 & 76.47 & 84.59 & 80.93 & 75.24 & 
				84.47 & 80.00 & 73.84 & 85.08 & 80.83 & 73.86 & 81.48  \\

				\textbf{UREV} & 81.98 & 88.09 & 
				90.07 & 87.38 & 89.63 & 81.10 & 78.51 & 81.46 & 79.50 & 
				77.47 & 84.59 & 79.88 & 74.69 & 85.39 & 80.31 & 76.07 & 
				83.50 & 81.41 & 73.46 & 84.31 & 82.42 & 74.67 & 81.63  \\	
				
				\textbf{UGCPREV} & 81.77 & 88.69 & 90.01 & 87.32 &
				\textbf{89.79} & 80.70 & \textbf{81.60} & 82.10 & 81.43 & 79.80 &
				82.85 & 81.06 & 76.88 & 84.91 & 81.81 & 76.37 &
				84.00 & 81.40 & 74.76 & 84.96 & 80.99 & 74.05 &
				82.15  \\
				
				\textbf{FetusMap} & 83.21 & 87.20 & 89.55 & 88.00 &
				89.73 & 81.10 & 80.59 & 80.91 & 78.60 & 80.20 &
				81.87 & 75.18 & 69.34 & 83.09 & 77.39 & 69.19 &
				83.66 & 77.19 & 68.33 & 84.32 & 78.30 & 69.54 &
				79.84  \\
				
				\textbf{FetusMapV2} & \textbf{83.51} & \textbf{88.97} & \textbf{90.53} & 
				\textbf{88.23} & 89.54 & \textbf{83.24} & 81.07 & 82.03 & \textbf{81.49} &
				\textbf{81.17} & 84.04 & 82.03 & 76.05 & \textbf{85.45} & 82.49 &
				\textbf{78.08} & \textbf{86.08} & \textbf{83.15} & \textbf{75.71} & \textbf{86.98} & \textbf{84.78} &
				\textbf{76.35} & \textbf{83.23}  \\
				
				\bottomrule[1.5pt]
		\end{tabular}}
		\label{tab:all_methods_auc}
	\end{center}
\end{table*}

For step one, we choose five landmarks from head and trunk (i.e., 2, 3, 5, 7, 9) with the highest detection accuracy across the validation set to generate the transformation matrix ($T_{a}$) for registration, as shown in Fig.~\ref{fig:framework}.
For step two, we retrieved the top-K pose with the lowest registration errors as the set of aligned pose $D_{t}$, the selection formula is as follows:
\begin{equation}
D_{t} = \mathop{\arg\min}_{D'\subset S,|D'|=K} \sum_{a=1}^{N} \sum_{j \in M}||T_{a}^{t} \times P_{a}^{j} - P_{t}^{j}||_{2},
\label{Equ.SSL1}
\end{equation}
where $K = 8$, and $M=\{2,3,5,7,9\}$ means the landmarks used for registration.
For unseen testing fetus volume, the detector predicts the $22$-channel landmark heatmaps {($H(P_{t}^{i}), i=1,2,\cdots,22$)} and original 3D pose estimation $P_t$. Each pose $P_a$ in the library $S$ is then aligned to $P_t$ via a transformation $T_{a}^{t}$. 
Finally, the pseudo label $P_p$ is generated from $P_a$ and $P_t$ (see Fig.~\ref{fig:framework}):

\begin{equation}
P_{p}^{i} = \frac{1}{2}(P_{t}^{i} + \dfrac{\sum_{a=1}^{K}T_{a}^{t} \times P_{a}^{i}}{K}),  i\in \left\lbrace 1, 4, 6, 8, 10 \right\rbrace,
\label{Equ.SSL2}
\end{equation}

where $i$ represents the index of ambiguous landmarks that needs to be optimized. The pseudo label synthesized by combining the model prediction with the shape knowledge of the pose library will be used as the supervision to optimize the prediction results online. The loss function is as follows: 
{
\begin{equation}
L_{ssl} = \frac{1}{5}\sum (H(P_{p}^{i}) - H(P_{t}^{i}))^2,  i\in \left\lbrace 1, 4, 6, 8, 10 \right\rbrace.
\label{Equ.SSL_loss}
\end{equation}
}

In practice, each case will have its own specific label proxy.
The effectiveness of SSL will be elaborated in Section \ref{sec:experiment}.\par

\section{Experiments}
\label{sec:experiment}

\subsection{Materials and implementation}
In this work, the fetus dataset consists of US volumes from 1000 pregnant volunteers with singletons at gestational ages ranging from 10 to 15 weeks. The dataset has an average size of $460\times313\times312$, with a resolution of $0.5\times0.5\times0.5 mm^3$ or $0.3\times0.3\times0.3 mm^3$ in different systems. Approved by local IRB, all volumes were anonymized and obtained by experts using Mindary, Philips and GE systems. The fetuses were permitted to be in any pose during data acquisition. Using the \textbf{Pair} annotation software package \citep{liang2022sketch}, three experts with over five years of experience manually annotated 22 landmarks (refer to Fig.~\ref{fig:Instance}(a)). {All three experts have been trained by senior sonographer to annotate and their annotated samples have no overlap.} The dataset was randomly divided into three sets of 600, 200, and 200 volumes for training, validation, and testing, respectively. The augmentation strategy used in this study involved a random $90^\circ$ rotation and scaling.\par

We implemented our method in the PyTorch \citep{paszke2019pytorch} framework, using a standard PC outfitted with a  12 GB NVIDIA GTX 2080Ti GPU. To ensure a fair and unambiguous comparison, we maintained a consistent input size limit for all methods, which was the maximum size possible within our memory constraints. Therefore, the original US volume was downscaled to $176\times112\times112$. However, our proposed memory-efficient architecture enabled us to increase the fixed input size to $208\times128\times144$. During the training stage, we utilized the Adam optimizer \citep{kingma2014adam} (batch size = $1$, initial learning rate = $10^{-4}$, moment term = $0.5$, and epoch = $100$) to update the weight in the network. {The batch normalization used in our model doesn't track running statistics. It always employs the current batch statistics during both training and evaluation modes. To implement this, we used $nn.BatchNorm3d(num\_channels, track\_running\_stats=False)$ in our PyTorch code.} During the testing with the SSL, we used an initial learning rate of $5\times 10^{-4}$ and operated the pre-trained detector for 8 iterations for each testing case.\par

The performances of different landmark localization methods were evaluated based on two key factors: localization accuracy and memory consumption. The localization accuracy was evaluated by the Euclidean Distance (ED) error ($mm$) between the predicted landmark and the ground truth and the AUC ($\%$) of 22 landmarks among the different methods, where the AUC is the area under the Percentage of Correct Keypoint (PCK) curve. Memory consumption was evaluated by comparing the resource occupancy and input resolution across the different methods. All experimental results were based on the average of the two-fold cross-validation, and the figures of PCK curves were derived from the first fold.\par

{{To evaluate the effectiveness of FetusMapV2, we conducted an extensive series of comparative experiments, including the exploration of different network architectures (UNet~\citep{isensee2018no}, UGAN~\citep{chen2017adversarial}, URNN~\citep{liu2019feature}), diverse GPU management schemes (UGCP, UREV, UGCPREV) under the same memory limit, different input sizes and input modes (patch, whole volume) and the consideration of \emph{Pair Loss} and SSL utilization, among other factors. The details of different methods are presented in Table \ref{table_METHOD_detail}. The network structure for all methods is UNet. Specifically, UPatch indicates the use of patch as the input mode. UPSSL refers to UNet equipped with our proposed \emph{Pair Loss} and SSL. UGCP, UREV, and UGCPREV refer to using gradient checkpoints, reversible layers, and our proposed memory-efficient architecture, respectively. In addition, we also compared FetusMapV2 with the version before upgrading FetusMap ~\citep{yang2019fetusmap}.}}

\begin{figure*}[!h]
	\centering
	\includegraphics[width=1.0 \linewidth]{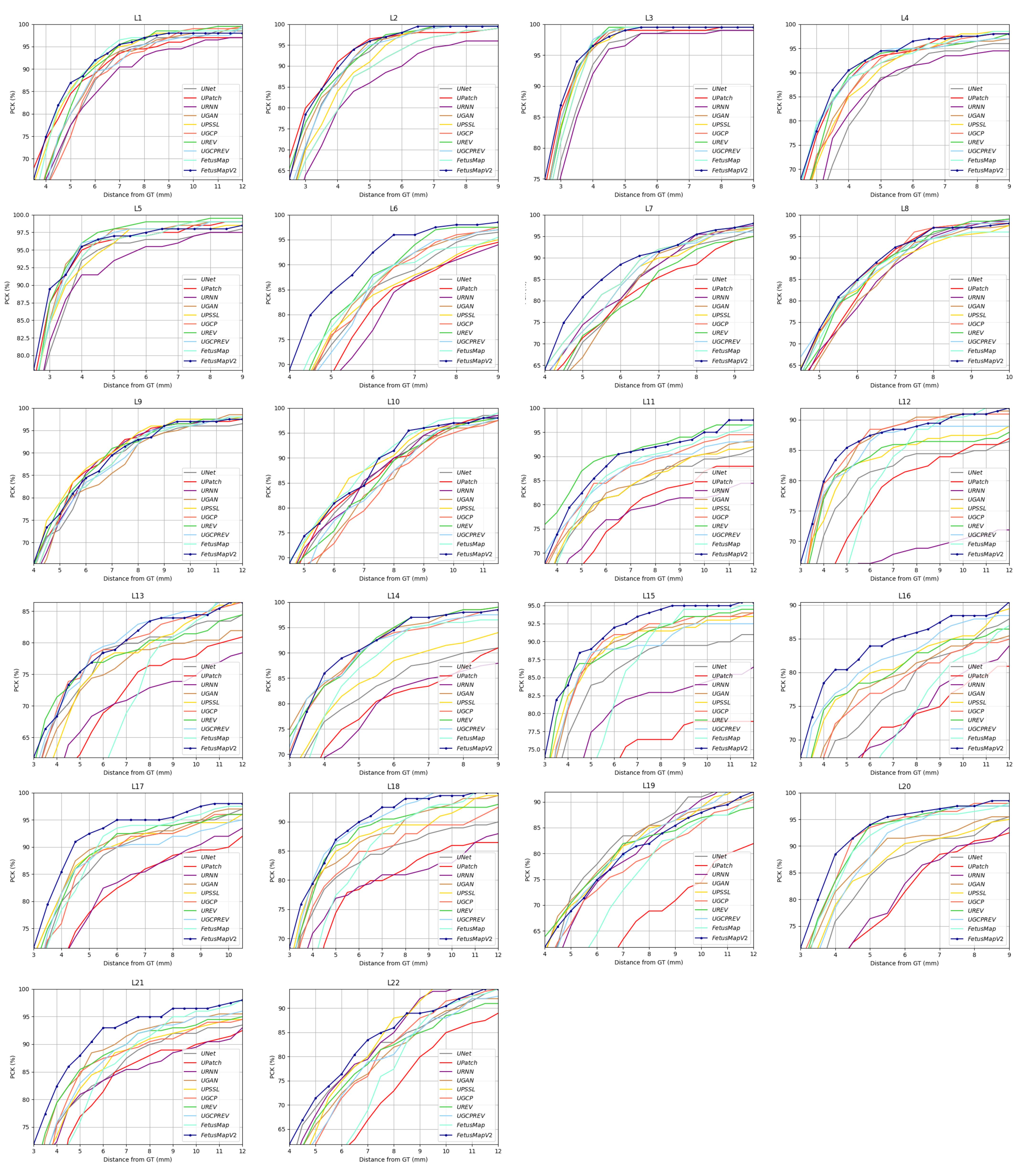}
	\caption{{PCK curves of 22 landmarks of different methods.}}
	\label{fig:pck}
	\vspace{-0.3cm}	
\end{figure*}

\begin{table*}[tb!]\footnotesize
        \setlength{\abovecaptionskip}{0.2cm}
	\centering
	\renewcommand\arraystretch{1}
	\begin{center}
		\caption{{{Comparison of different methods on input size, resource consumption, parameters, training time and performance. UNet-L refers to the UNet method with larger input sizes, and Average Time refers to the average time consumed per volume during training.}}}
            \resizebox{\textwidth}{!}{
			\begin{tabular}{c|c|c|c|c|c|c}
                    \hline
				\multirow{2}{2cm }{ \centering \textbf{Methods}}& \multirow{2}{2.2cm}{\centering \textbf{{Input Size}}}  & {\multirow{2}{1.5cm}{\centering \textbf{Memory \\ (\textit{GB})}}} &{\multirow{2}{1.6cm}{\centering \textbf{Parameters \\ (\textit{MB})}}}& {\multirow{2}{2cm }{\centering \textbf{Average Time \\ (\textit{s})}}}& {\multirow{2}{2.2cm}{\centering \textbf{Average ED error \\ (\textit{mm}) $\downarrow$ }}} &{\multirow{2}{1.8cm}{\centering \textbf{Average AUC \\(\textit{$\%$}) $\uparrow$}}}\\ 
                & & & & & & \\
                 \hline
			{\textbf{UPatch}} & {[176, 112, 112]} & {7.91 (100$\%$)}& {147.98} &{40.20} &{5.07} &{77.35}  \\
				
               \textbf{UNet} & [176, 112, 112] & {7.91 (100$\%$)}& {147.98} &1.47 & {4.14} & {80.44}\\
               
               {\textbf{UNet-L} }& {[256, 160, 160] }& {23.81 (294$\%$)}& {147.98} & {--} & {3.54}& {82.61} \\
    
				\textbf{UGCP} & [176, 112, 112] & {5.95 (75$\%$)} & {147.98} & 1.82 & {4.16}& {80.40}\\
								
				\textbf{UREV} & [176, 112, 112] & {5.21 (66$\%$)}&{148.01} & 2.16 & {4.12 }& {80.47}\\
								
				\textbf{UGCPREV} & [176, 112, 112] & {4.50(57$\%$)} &{147.99 }&{1.83}& {4.11}& {80.49}\\
								
				\hline
				\textbf{UGCP} & [192, 112, 128] & {7.36} & {147.98}& 2.31 & {3.90}& {81.48}\\
				\textbf{UREV} & [192, 128, 128] & {7.33} & {148.01}& 3.13 & {3.84}& {81.63}\\
				\textbf{UGCPREV} & [208, 128, 144] & {7.38} &{147.99} & 3.23 & {3.73} & {82.15}\\

             \hline
		\end{tabular}}
		\label{tab:memory1}
	\end{center}
\end{table*}

\begin{figure*}[htb]
	\centering
	\includegraphics[width=0.95 \linewidth]{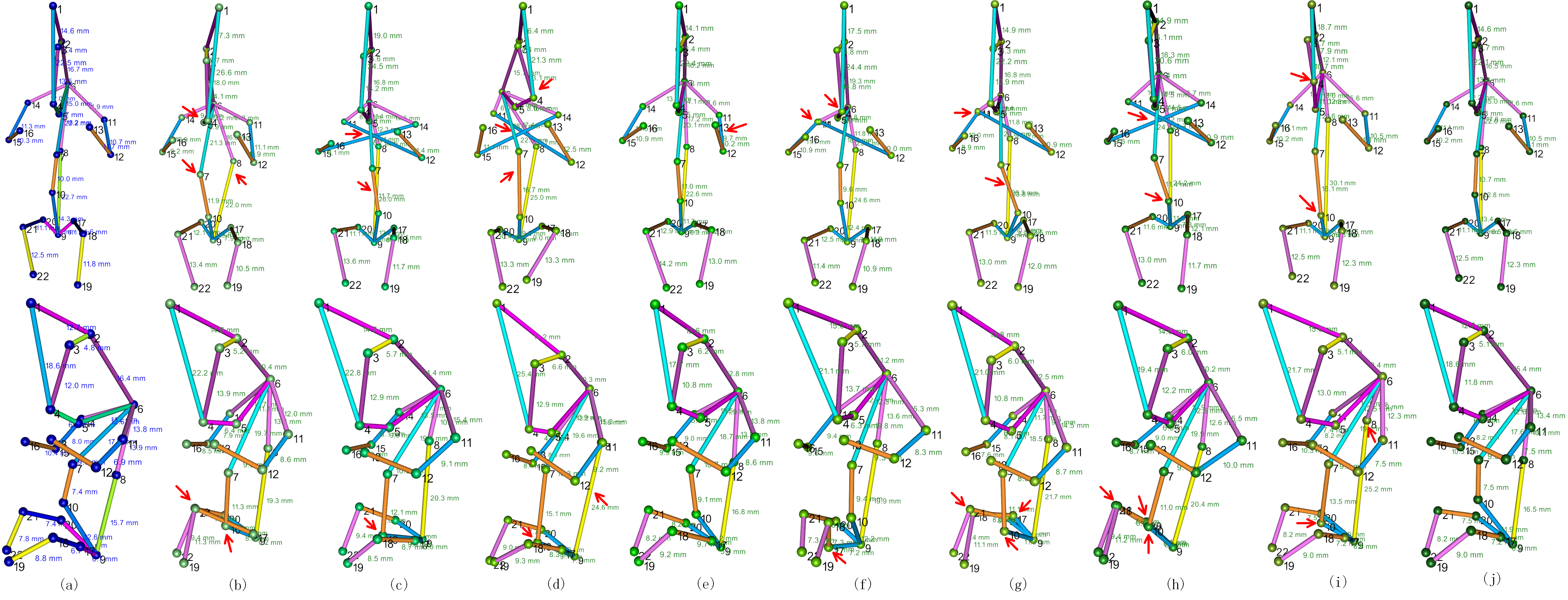}
	\caption{Results of different methods. From left to right: (a) \textbf{Ground Truth}. (b) UNet. (c) UGAN. (d) URNN. (e) UPSSL. (f) UGCP.(g) UREV. (h) UGCPREV. (i) FetusMapV2 w/o SSL. (j) \textbf{FetusMapV2}.}
	\label{fig:result}
	\vspace{-0.3cm}	
\end{figure*}

\subsection{Main results}
Table \ref{tab:all_methods_dis} summarizes the localization results for all 22 key landmarks regarding the  ED error. Compared with all other methods, our proposed method named FetusMapV2 achieved the most satisfactory results at 14 landmarks and had the lowest average ED error. This demonstrates the effectiveness of the three proposed optimization schemes. To further highlight the superiority of these optimization schemes, we compared the same optimization strategies from different perspectives. {{Comparing the performance of the UPatch and UNet methods in the first two rows of Table \ref{tab:all_methods_dis}, it is evident that whole volume input outperforms patch input, which can be attributed to more comprehensive global information provided by whole volume input. Regarding utilizing landmark information for further optimization, our proposed UPSSL method outperformed URNN and UGAN, particularly on the limb landmarks and the structural ambiguity landmarks. This verifies the effectiveness of our proposed \emph{Pair Loss} and SSL methods in simplifying the learning process for limb landmarks and effectively utilizing prior knowledge. Furthermore, it can be seen that UGCPREV achieved the lowest ED error compared to UGCP and UREV, primarily attributed to the higher resolution input facilitated by our proposed efficient memory management scheme.}} Moreover,  although our initial work, FetusMap, used the same gradient checkpoint technology as the UGCP, the performance of the limb landmarks was not as excellent as that of UGCP, as shown in Table \ref{tab:all_methods_dis}. We attribute this performance degradation to the increase in the variation of the fetal pose with the increasing cases, {{as shown in Fig. \ref{fig:Instance}(c), }}making it infeasible to use SSL on the landmarks of the limbs in FetusMap. Therefore, we applied the SSL to the landmarks of the head and trunk, whose positional relationship was relatively fixed. The experimental results demonstrate the effectiveness of our adjustment.\par

The PCK evaluates the distribution of the predicted landmarks around the ground truth. Table \ref{tab:all_methods_auc} further compares the AUC of the different methods. The results are consistent with those in Table \ref{tab:all_methods_dis}, with FetusMapV2 achieving the best results at 15 landmarks and having the highest mean AUC among all competitors. Notably, our three contributions (GCPREV, \emph{Pair Loss}, and SSL) were responsible for the best results for all landmarks. This demonstrates that each of our contributions significantly improves the basic network, and that combining them further enhances the performance. We further show the PCK curves for all landmarks from the different methods in Fig.~\ref{fig:pck}. FetusMapV2 achieved the best results on multiple landmarks and demonstrated significant improvements over other schemes such as $L_6$, $L_7$, $L_{16}$, $L_{17}$ and $L_{21}$. In particular, in the $L_{16}$ curve of Fig.~\ref{fig:pck}, we observe that higher-resolution inputs can alleviate the problem of landmark confusion (UGCPREV performs better than UPSSL), and combining GCPREV and \emph{Pair Loss} can achieve the best result. \par

\begin{figure*}[htb]
	\centering
	\includegraphics[width=0.95 \linewidth]{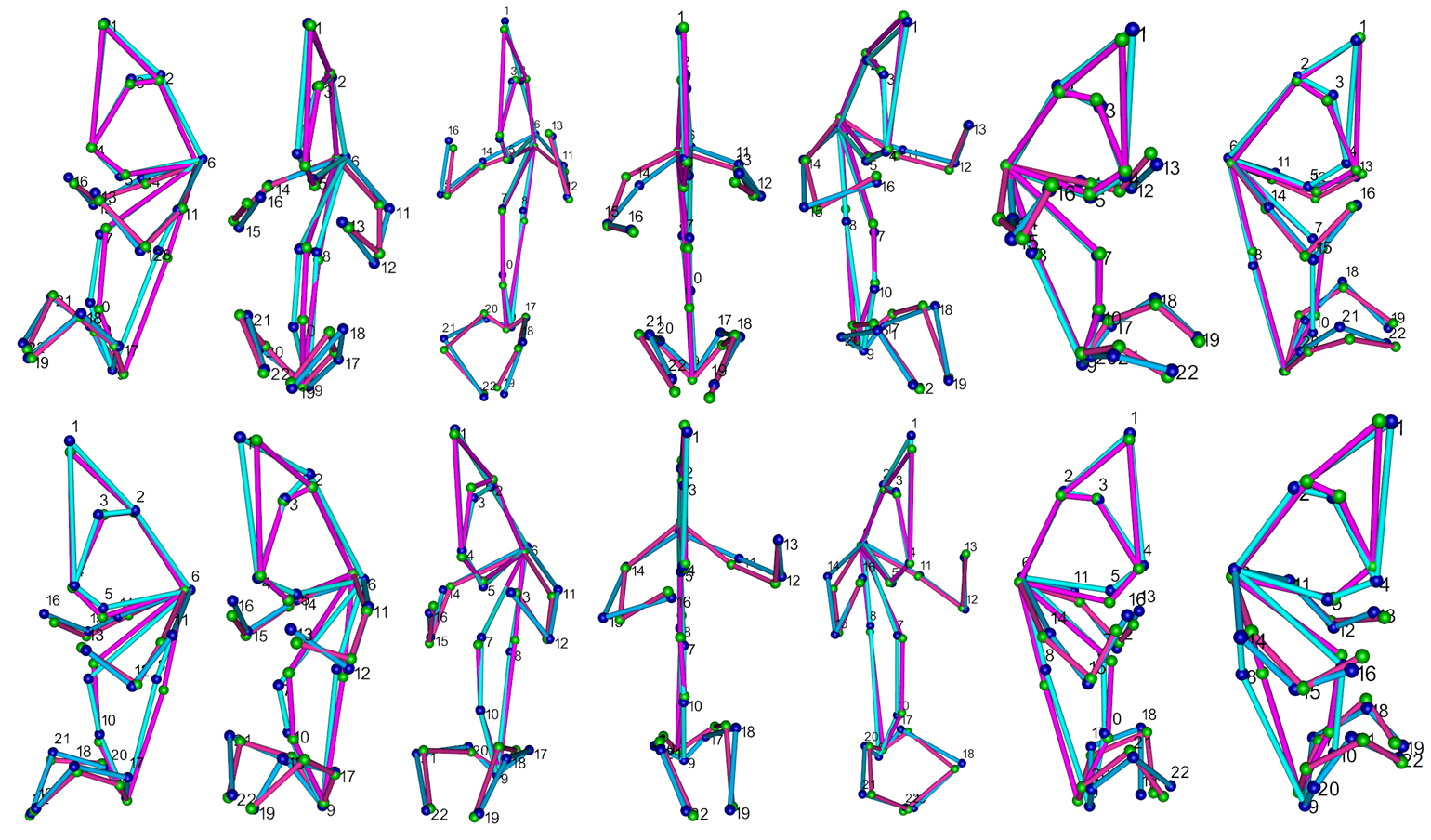}
	\caption{Visualisation of the predicted results of 14 fetal cases with different orientations. Blue landmarks and blue connections are the ground truth, and green landmarks and purple connections are our predictions.}
	\label{fig:more_result}
	\vspace{-0.3cm}	
\end{figure*}

Additionally, compared to the basic network, the proposed FetusMapV2 framework lowered the mean Euclidean distance (ED) error of the 22 landmarks from $4.14 mm$ to $3.46 mm$, and the mean AUC increased from $80.44\%$ to $83.23\%$ as shown in Table ~\ref{tab:all_methods_dis} and \ref{tab:all_methods_auc}. Furthermore, We discovered that FetusMapV2 used memory efficiently while obtaining better results. The outputs from the comparison of different methods are depicted in Fig. \ref{fig:result}. We found that the SSL and Pair-Loss can effectively correct the key landmarks of the limbs and torso (see Fig. \ref{fig:result}(h) and (j)), respectively, and the estimated pose of FetusMapV2 is the closest to the ground truth. More results of different poses for FetusMapV2 are shown in Fig. \ref{fig:more_result}.

\subsection{Memory consumption under different architectures}
{{Table \ref{tab:memory1} summarizes the comparison of different methods on input size, resource consumption, parameters, training time, and model performance. Memory consumption was calculated using the PyTorch memory profiling tool $torch.cuda.max\_memory\_allocated$, which computes the maximum GPU memory occupied by tensors, including the total memory of the model, optimizer, and activation. Note that UNet-L can only be trained in a 24GB NVIDIA RTX 4090 GPU due to the large memory it occupies. The statistical training time is not fair compared with other methods, so it was not listed in Table \ref{tab:memory1}. Comparing the second and third rows of Table \ref{tab:memory1}, larger input resolutions notably enhance landmark localization. But it's essential to highlight that this enhancement comes at a cost - four times resource consumption during training. Therefore, there is a pressing need for an efficient memory management scheme to alleviate the substantial memory demands associated with higher-resolution inputs.}} As shown in the first {{six}} rows of Table~\ref{tab:memory1}, our proposed method achieved a memory reduction of $43\%$ relative to the peak memory consumption of the vanilla run, outperforming previous methods. Our proposed scheme allows for larger input sizes(enlarged from $(176, 112, 112)$ to $(208, 128, 144)$) with the same memory usage, as shown in the last three rows of Table~\ref{tab:memory1}. Regarding the training time, although UGCPREV takes 1.24 times longer than UNet with the same input shape, it is comparable to the shortest time consumption of UGCP and UREV. This indicates that the optimization of these two schemes can be effectively combined. Furthermore, when comparing UGCPREV with UREV under the same memory limitation, the training time only increased by $3\%$ when the input resolution increased by 20\%. This indicates that the efficiency of our proposed memory management scheme. {In addition, compareing the results of ED and AUC of FetusMapV2 in Table~\ref{tab:all_methods_dis} and \ref{tab:all_methods_auc} with those of UNet-L in Table~\ref{tab:memory1}, FetusMapV2 achieves better performance to UNet-L while consuming much fewer resources.}\par

\subsection{Ablation experiments}
In this study, in addition to the memory-efficient architecture, we employ SSL and utilize \emph{Pair Loss} to refine different key landmarks separately. We conducted several ablations to evaluate the effectiveness of FetusMapV2, which are presented and discussed in detail in Tables~\ref{tab:ablation_memory}, \ref{tab:ablation_hc} and \ref{tab:ablation_ssl}.\par

\begin{table}[tb!]
	\centering
	\renewcommand\arraystretch{0.8}
	\begin{center}
		\caption{Experimental results of different methods under the same memory limitation.}
		\setlength{\tabcolsep}{1.2mm}{
			\begin{tabular}{c|c|c|c|c||c|c|c|c}
				\toprule[1.5pt]
				\multirow{3}{*}{\scriptsize\textbf{Landmarks}} & \multicolumn{4}{c||}{\footnotesize\textbf{AUC ($\%$) $\uparrow$ }} & \multicolumn{4}{c}{\footnotesize\textbf{ED ($mm$) $\downarrow$ }} \\
				\cline{2-9} 

				& \multirow{2}{*}{\tiny\textbf{UNet}} & \multirow{2}{*}{\tiny\textbf{UGCP}} & \multirow{2}{*}{\tiny\textbf{UREV}} & \multirow{2}{*}{\tiny\textbf{UGCPREV}} 
				& \multirow{2}{*}{\tiny\textbf{UNet}} & \multirow{2}{*}{\tiny\textbf{UGCP}} & \multirow{2}{*}{\tiny\textbf{UREV}} & \multirow{2}{*}{\tiny\textbf{UGCPREV}}   \\
				&                                     &                                      &                                    &  
				&                                      &                                     &                                    &   \\
				
				\hline
				\scriptsize{\textbf{$L_{1}$}} & \scriptsize{80.08} &\scriptsize{81.30} & \scriptsize\textbf{81.98} & \scriptsize{81.77} & 
				\scriptsize{4.07} & \scriptsize{4.07} & \scriptsize\textbf{3.62} & \scriptsize{3.85} \\
				
				\scriptsize{\textbf{$L_{2}$}} &\scriptsize{87.69} &\scriptsize{87.20} & \scriptsize{88.09} & \scriptsize\textbf{88.69} & 
				\scriptsize{2.46} & \scriptsize{2.56} & \scriptsize{2.38} & \scriptsize\textbf{2.26} \\
				
				\scriptsize{\textbf{$L_{3}$}} &\scriptsize{88.77} &\scriptsize{89.55} & \scriptsize\textbf{90.07} & \scriptsize{90.01} & 
				\scriptsize{2.30} & \scriptsize{2.10} & \scriptsize\textbf{1.98} & \scriptsize{2.01} \\
				
				\scriptsize{\textbf{$L_{4}$}}& \scriptsize{84.46} &\scriptsize\textbf{87.64} & \scriptsize{87.38} & \scriptsize{87.32} & 
				\scriptsize{3.14} & \scriptsize\textbf{2.47} & \scriptsize{2.56} & \scriptsize{2.56} \\
				
				\scriptsize{\textbf{$L_{5}$}} & \scriptsize{89.20} &\scriptsize{89.74} & \scriptsize{89.63} & \scriptsize\textbf{89.79} & 
				\scriptsize{2.16} & \scriptsize{2.04} & \scriptsize{2.06} & \scriptsize\textbf{2.02} \\
				
				\scriptsize{\textbf{$L_{6}$}} & \scriptsize{80.59} &\scriptsize{81.00} & \scriptsize\textbf{81.10} & \scriptsize{80.70} & 
				\scriptsize{3.97} & \scriptsize\textbf{3.83} & \scriptsize{3.84} & \scriptsize{3.90} \\
				
				\scriptsize{\textbf{$L_{7}$}} & \scriptsize{79.78} &\scriptsize{80.59} & \scriptsize{78.51} & \scriptsize\textbf{81.60} & 
				\scriptsize{4.04} & \scriptsize{3.88} & \scriptsize{4.29} & \scriptsize\textbf{3.68} \\
				
				\scriptsize{\textbf{$L_{8}$}} & \scriptsize\textbf{82.45} &\scriptsize{81.39} & \scriptsize{81.46} & \scriptsize{82.10} & 
				\scriptsize\textbf{3.51} & \scriptsize{3.80} & \scriptsize{3.73} & \scriptsize{3.67} \\
				
				\scriptsize{\textbf{$L_{9}$}} &\scriptsize{79.02} &\scriptsize{78.60} & \scriptsize{79.50} & \scriptsize\textbf{81.43} & 
				\scriptsize{4.93} & \scriptsize{4.47} & \scriptsize{4.16} & \scriptsize\textbf{3.73} \\
				
				\scriptsize{\textbf{$L_{10}$}} &\scriptsize{79.39} &\scriptsize{77.34} & \scriptsize{77.47} & \scriptsize\textbf{79.80} & 
				\scriptsize{4.22} & \scriptsize{4.53} & \scriptsize{4.50} & \scriptsize\textbf{4.14} \\
				
				\scriptsize{\textbf{$L_{11}$}} & \scriptsize{78.23} &\scriptsize{82.72} & \scriptsize\textbf{84.59} & \scriptsize{82.85} & 
				\scriptsize{4.59} & \scriptsize{3.65} & \scriptsize\textbf{3.17} & \scriptsize{3.62} \\
				
				\scriptsize{\textbf{$L_{12}$}} &\scriptsize{77.46} &\scriptsize{80.20} & \scriptsize{79.88} & \scriptsize\textbf{81.06} & 
				\scriptsize{5.03} & \scriptsize{4.50} & \scriptsize{4.57} & \scriptsize\textbf{4.20} \\
				
				\scriptsize{\textbf{$L_{13}$}} &\scriptsize{74.53} &\scriptsize{76.47} & \scriptsize{74.69} & \scriptsize\textbf{76.88} & 
				\scriptsize{5.59} & \scriptsize{5.06} & \scriptsize{5.50} & \scriptsize\textbf{4.87} \\
				
				\scriptsize{\textbf{$L_{14}$}}& \scriptsize{79.91} &\scriptsize{84.59} & \scriptsize\textbf{85.39} & \scriptsize{84.91} & 
				\scriptsize{4.04} & \scriptsize{3.13} & \scriptsize\textbf{2.95} & \scriptsize{3.13} \\
				
				\scriptsize{\textbf{$L_{15}$}} & \scriptsize{78.46} &\scriptsize{80.93} & \scriptsize{80.31} & \scriptsize\textbf{81.81} & 
				\scriptsize{4.80} & \scriptsize{4.21} & \scriptsize{4.29} & \scriptsize\textbf{3.80} \\
				
				\scriptsize{\textbf{$L_{16}$}} & \scriptsize{74.19} &\scriptsize{75.24} & \scriptsize{76.07} & \scriptsize\textbf{76.37} & 
				\scriptsize{5.62} & \scriptsize{5.78} & \scriptsize{5.21} & \scriptsize\textbf{5.14} \\
				
				\scriptsize{\textbf{$L_{17}$}} & \scriptsize{83.24} &\scriptsize\textbf{84.47} & \scriptsize{83.50} & \scriptsize{84.00} & 
				\scriptsize{3.59} & \scriptsize\textbf{3.16} & \scriptsize{3.44} & \scriptsize{3.29} \\
				
				\scriptsize{\textbf{$L_{18}$}} & \scriptsize{79.24} &\scriptsize{80.00} & \scriptsize\textbf{81.41} & \scriptsize{81.40} & 
				\scriptsize{4.27} & \scriptsize{4.22} & \scriptsize{3.95} & \scriptsize\textbf{3.90} \\
				
				\scriptsize{\textbf{$L_{19}$}} &\scriptsize\textbf{75.59} &\scriptsize{73.84} & \scriptsize{73.46} & \scriptsize{74.76} & 
				\scriptsize\textbf{5.39} & \scriptsize{5.77} & \scriptsize{5.76} & \scriptsize{5.50} \\
				
				\scriptsize{\textbf{$L_{20}$}} &\scriptsize{82.66} &\scriptsize\textbf{85.08} & \scriptsize{84.31} & \scriptsize{84.96} & 
				\scriptsize{3.71} & \scriptsize\textbf{3.02} & \scriptsize{3.19} & \scriptsize{3.20} \\
				
				\scriptsize{\textbf{$L_{21}$}} &\scriptsize{80.31} &\scriptsize{80.83} & \scriptsize\textbf{82.42} & \scriptsize{80.99} & 
				\scriptsize{4.14} & \scriptsize{3.96} & \scriptsize\textbf{3.64} & \scriptsize{4.01} \\
				
				\scriptsize{\textbf{$L_{22}$}} &\scriptsize{74.42} &\scriptsize{73.86} & \scriptsize\textbf{74.67} & \scriptsize{74.05} & 
				\scriptsize{5.55} & \scriptsize\textbf{5.55} & \scriptsize{5.79} & \scriptsize{5.66} \\
				
				\hline 
				
				\scriptsize{\textbf{$mean$}} & \scriptsize{80.44} &\scriptsize{81.48} & \scriptsize{81.63} & \scriptsize\textbf{82.15} & 
				\scriptsize{4.14} & \scriptsize{3.90} & \scriptsize{3.84} & \scriptsize\textbf{3.73} \\	
				
				\bottomrule[1.5pt]

		\end{tabular}}
		\label{tab:ablation_memory}
	\end{center}
\end{table}

\begin{table}[tb!]
	\centering
	\renewcommand\arraystretch{0.8}
	\begin{center}
		\caption{ Ablation experiment results on \emph{Pair Loss}.}
		\setlength{\tabcolsep}{0.8mm}{
			\begin{tabular}{c|c|c|c|c||c|c|c|c}
				\toprule[1.5pt]
				\multirow{3}{*}{\scriptsize\textbf{Landmarks}} & \multicolumn{4}{c||}{\footnotesize\textbf{AUC ($\%$) $\uparrow$ }} & \multicolumn{4}{c}{\footnotesize\textbf{ED ($mm$) $\downarrow$ }} \\
				\cline{2-9} 
				& \multirow{2}{*}{\tiny\textbf{UNet}} &  \multirow{2}{*}{\tiny\textbf{UP}} &  \tiny\textbf{FetusMapV2}  & \multirow{2}{*}{\tiny\textbf{FetusMapV2}} 
				& \multirow{2}{*}{\tiny\textbf{UNet}} & \multirow{2}{*}{\tiny\textbf{UP}} &  \tiny\textbf{FetusMapV2}  & \multirow{2}{*}{\tiny\textbf{FetusMapV2}} \\
				&  &  & \tiny\textbf{-w/o P} &  &  &  & \tiny\textbf{-w/o P} & \\
				\hline
				\scriptsize{\textbf{$L_{11}$}} & \scriptsize{78.23} &\scriptsize{81.53} & \scriptsize{82.85} & \scriptsize{\textbf{84.04}} & 
				\scriptsize{4.59} & \scriptsize{3.81} & \scriptsize{3.62} & \scriptsize{\textbf{3.20}} \\
				
				\scriptsize{\textbf{$L_{12}$}} &\scriptsize{77.46} &\scriptsize{80.70} & \scriptsize{81.06} & \scriptsize{\textbf{82.03}} & 
				\scriptsize{5.03} & \scriptsize{4.16} & \scriptsize{4.20} & \scriptsize{\textbf{3.88}} \\
				
				\scriptsize{\textbf{$L_{13}$}} &\scriptsize{74.53} &\scriptsize{76.65} & \scriptsize{\textbf{76.88}} & \scriptsize{76.05} & 
				\scriptsize{5.59} & \scriptsize{4.91} & \scriptsize{\textbf{4.87}} & \scriptsize{5.09} \\
				
				\scriptsize{\textbf{$L_{14}$}}& \scriptsize{79.91} &\scriptsize{83.02} & \scriptsize{84.91} & \scriptsize{\textbf{85.45}} & 
				\scriptsize{4.04} & \scriptsize{3.45} & \scriptsize{3.13} & \scriptsize{\textbf{2.91}} \\
				
				\scriptsize{\textbf{$L_{15}$}} & \scriptsize{78.46} &\scriptsize{\textbf{83.25}} & \scriptsize{81.81} & \scriptsize{82.49} & 
				\scriptsize{4.80} & \scriptsize{\textbf{3.55}} & \scriptsize{3.80} & \scriptsize{3.77} \\
				
				\scriptsize{\textbf{$L_{16}$}} & \scriptsize{74.19} &\scriptsize{76.93} & \scriptsize{76.37} & \scriptsize{\textbf{78.08}} & 
				\scriptsize{5.62} & \scriptsize{5.22} & \scriptsize{5.14} & \scriptsize{\textbf{4.59}} \\
				
				\scriptsize{\textbf{$L_{17}$}} & \scriptsize{83.24} &\scriptsize{84.57} & \scriptsize{84.00} & \scriptsize{\textbf{86.08}} & 
				\scriptsize{3.59} &  \scriptsize{3.09} & \scriptsize{3.29} & \scriptsize{\textbf{2.81}} \\
				
				\scriptsize{\textbf{$L_{18}$}} & \scriptsize{79.24} &\scriptsize{81.71} & \scriptsize{81.40} & \scriptsize{\textbf{83.15}} & 
				\scriptsize{4.27} &  \scriptsize{3.82} & \scriptsize{3.90} & \scriptsize{\textbf{3.41}} \\
				
				\scriptsize{\textbf{$L_{19}$}} &\scriptsize{75.59} &\scriptsize{75.27} & \scriptsize{74.76} & \scriptsize{\textbf{75.71}} & 
				\scriptsize{\textbf{5.39}} &  \scriptsize{5.51} & \scriptsize{5.50} & \scriptsize{5.46} \\
				
				\scriptsize{\textbf{$L_{20}$}} &\scriptsize{82.66} &\scriptsize{84.31} & \scriptsize{84.96} & \scriptsize{\textbf{86.98}} & 
				\scriptsize{3.71} &  \scriptsize{3.14} & \scriptsize{3.20} & \scriptsize{\textbf{2.66}} \\
				
				\scriptsize{\textbf{$L_{21}$}} &\scriptsize{80.31} &\scriptsize{81.41} & \scriptsize{80.99} & \scriptsize{\textbf{84.78}} & 
				\scriptsize{4.14} &  \scriptsize{3.97} & \scriptsize{4.01} & \scriptsize{\textbf{3.05}} \\
				
				\scriptsize{\textbf{$L_{22}$}} &\scriptsize{74.42} &\scriptsize{74.89} & \scriptsize{74.05} & \scriptsize{\textbf{76.35}} & 
				\scriptsize{5.55} &  \scriptsize{5.47} & \scriptsize{5.66} & \scriptsize{\textbf{5.13}} \\
				
				\hline
				
				\scriptsize{\textbf{$mean$}} & \scriptsize{78.19} &\scriptsize{80.35} & \scriptsize{80.34} & \scriptsize{\textbf{81.76}} & 
				\scriptsize{4.69} & \scriptsize{4.17} & \scriptsize{4.19} & \scriptsize{\textbf{3.83}} \\
				\bottomrule[1.5pt]
				
		\end{tabular}}
		\label{tab:ablation_hc}
	\end{center}
\end{table}

\begin{table}[tb!]
	\centering
	\renewcommand\arraystretch{0.8}
	\begin{center}
		\caption{Ablation experiment results on SSL.}
		\setlength{\tabcolsep}{0.8mm}{
			\begin{tabular}{c|c|c|c|c||c|c|c|c}
				\toprule[1.5pt]
				\multirow{3}{*}{\scriptsize\textbf{Landmarks}} & \multicolumn{4}{c||}{\footnotesize\textbf{AUC ($\%$) $\uparrow$ }} & \multicolumn{4}{c}{\footnotesize\textbf{ED ($mm$) $\downarrow$ }} \\
				\cline{2-9} 
				& \multirow{2}{*}{\tiny\textbf{UNet}} & \multirow{2}{*}{\tiny\textbf{USSL}} & \tiny\textbf{FetusMapV2} & \multirow{2}{*}{\tiny\textbf{FetusMapV2}} 
				& \multirow{2}{*}{\tiny\textbf{UNet}} & \multirow{2}{*}{\tiny\textbf{USSL}} & \tiny\textbf{FetusMapV2} & \multirow{2}{*}{\tiny\textbf{FetusMapV2}} \\
				
				&  &  & \tiny\textbf{-w/o SSL} &  &  &  & \tiny\textbf{-w/o SSL} &  \\
				\hline
				
				\scriptsize{\textbf{$L_{1}$}} & \scriptsize{80.08} &\scriptsize{81.57} & \scriptsize{81.62} & \scriptsize{\textbf{83.51}} & 
				\scriptsize{4.07} & \scriptsize{3.69} & \scriptsize{3.77} & \scriptsize{\textbf{3.32}} \\
				
				\scriptsize{\textbf{$L_{4}$}} &\scriptsize{84.46} &\scriptsize{85.97} & \scriptsize{87.95} & \scriptsize{\textbf{88.23}} & 
				\scriptsize{3.14} & \scriptsize{2.81} & \scriptsize{2.48} & \scriptsize{\textbf{2.37}} \\
				
				\scriptsize{\textbf{$L_{6}$}} &\scriptsize{80.59} &\scriptsize{80.91} & \scriptsize{81.18} & \scriptsize{\textbf{83.24}} & 
				\scriptsize{3.97} & \scriptsize{3.82} & \scriptsize{3.79} & \scriptsize{\textbf{3.35}} \\
				
				\scriptsize{\textbf{$L_{8}$}}& \scriptsize{\textbf{82.45}} &\scriptsize{81.38} & \scriptsize{81.12} & \scriptsize{82.03} & 
				\scriptsize{\textbf{3.51}} & \scriptsize{3.73} & \scriptsize{3.80} & \scriptsize{3.60} \\
				
				\scriptsize{\textbf{$L_{10}$}} & \scriptsize{79.39} &\scriptsize{79.88} & \scriptsize{80.30} & \scriptsize{\textbf{81.17}} & 
				\scriptsize{4.22} & \scriptsize{4.13} & \scriptsize{3.94} & \scriptsize{\textbf{3.75}} \\
				
				\hline
				
				\scriptsize{\textbf{$mean$}} & \scriptsize{81.39} &\scriptsize{81.94} & \scriptsize{82.43} & \scriptsize{\textbf{83.64}} & \scriptsize{3.78} &
				\scriptsize{3.64} & \scriptsize{3.56} & \scriptsize{\textbf{3.28}} \\
				\bottomrule[1.5pt]
				
		\end{tabular}}
		\label{tab:ablation_ssl}
	\end{center}
\end{table}

As indicated in Table \ref{tab:ablation_memory}, under the same memory occupation, increasing memory utilization through a higher input resolution led to improved landmark localization results. The 22 landmarks’ mean AUC and ED error improved $1.74 \%$ and $0.41 mm$, respectively. In particular, key landmarks corresponding to these ambiguous structures, such as $L_2$, $L_5$, $L_7$, $L_9$, and $L_{10}$, showed significant improvements. Furthermore, larger input resolutions also produce considerable improvements in the localization of the elbows and wrists ($L_{12}, L_{13}, L_{15}$, and $L_{16}$), which have a wide range of movements. We believe that these improvements were due to better feature perception with higher-resolution inputs, which further highlights the effectiveness of our memory reduction strategy. In addition, our proposed approach is flexible and can be applied to other models.\par

To address the confusion issues of symmetrical and similar landmarks, we designed \emph{Pair Loss}. As shown in Table~\ref{tab:ablation_hc}, the result of $L_{11}-L_{22}$ is significantly improved when using \emph{Pair Loss}. Comparing UNet and UP (UNet with \emph{Pair Loss}), we observed improvements in 11 out of 12 key landmarks in terms of the mean AUC and ED. The same improvement was observed when comparing FetusMapV2 w/o P (FetusMapV2 without using \emph{Pair Loss}) and FetusMapV2. We attribute the improvement in these landmarks to the fact that \emph{Pair Loss} reduces the task complexity to some extent. In contrast, the traditional heatmap regression method implies the classification of landmark categories. By separating this implicit task during optimization, \emph{Pair Loss} yielded significantly improved results.\par

For a more detailed comparison of the experiments, refer to the last row of Table \ref{tab:ablation_hc}. In the basic network, the mean AUC at 12 landmarks increased from $78.19\%$ to $80.35\%$, and the mean ED error decreases from $4.69mm$ to $4.17mm$. Additionally, compared to FetusMapV2 w/o P, FetusMapV2 demonstrated an increase in the mean AUC at 12 landmarks by $1.52\%$ and a decrease in the mean ED error by $0.36mm$. This result indicates that separating the hidden task of distinguishing the orientations can effectively alleviate the confusion issues. The visualized results of \emph{Pair Loss} are depicted in Fig.~\ref{fig:result}(e), (i), and (j).\par

Table~\ref{tab:ablation_ssl} presents an evaluation of our proposed SSL framework. To observe the improvement attained by SSL intuitively, we should analyze its performance on five landmarks $\left\lbrace1, 4, 6, 8, 10 \right\rbrace$. For this ablation experiment, we compared the results under the two baselines: UNet and FetusMapV2 without using the SSL (FetusMapV2 w/o SSL). Both baselines have been improved using the SSL refinement framework. SSL improves the mean AUC at five landmarks by $2.24\%$ and decreases the mean ED error by $0.6mm$ over the basic network. Furthermore, for FetusMapV2 w/o SSL, SSL can also improve the mean AUC at five landmarks by $1.24\%$ and decreases the mean ED error by $0.28mm$, with much of the gain attained at $L_1$ and $L_4$. By comparing (b) and (e), and (i) and (j) of Fig. \ref{fig:result}, we can get a more intuitive sense. Comparing Fig.~\ref{fig:result}(i) with  Fig.~\ref{fig:result} (j) in the first row, $L_4$ and $L_{10}$ have been improved significantly. Although the visualized result of FetusMapV2 w/o SSL as shown in Fig. \ref{fig:result}(i) is comparable among other competitors, this online refinement is necessary for more accurate applications.\par

\begin{table}[!t]
\centering
\renewcommand\arraystretch{0.8}
\begin{center}
\caption{The MAE of experiments on the FetusMapV2 applications.}
\setlength{\tabcolsep}{3.5mm}{
\begin{tabular}{c|c|c|c}
\toprule[1.5pt]  
\multicolumn{2}{c|}{\footnotesize\textbf{Applications}} & \footnotesize\textbf{UNet} & \footnotesize\textbf{FetusMapV2}  \\
\hline
\footnotesize\textbf{Midsagittal}& \footnotesize\textbf{Distance ($mm$)} & \scriptsize{22.50} & \scriptsize{20.00} \\ 
\cline{2-4} 
\footnotesize\textbf{Plane}& \footnotesize\textbf{Angle ($^\circ$)} & \scriptsize{9.55} & \scriptsize{8.34} \\
\hline 
\multicolumn{2}{c|}{\footnotesize\textbf{CRL ($mm$)}} & \scriptsize{9.13} & \scriptsize{6.28} \\
\hline 
\multicolumn{2}{c|}{\footnotesize\textbf{Left Humerus ($mm$)}} & \scriptsize{11.00} & \scriptsize{7.85} \\
\hline 
\multicolumn{2}{c|}{\footnotesize\textbf{Right Humerus ($mm$)}} & \scriptsize{9.08} & \scriptsize{7.53} \\
\hline 
\multicolumn{2}{c|}{\footnotesize\textbf{Left Femur ($mm$)}} & \scriptsize{8.43} & \scriptsize{6.95} \\
\hline 
\multicolumn{2}{c|}{\footnotesize\textbf{Right Femur ($mm$)}} & \scriptsize{8.16} & \scriptsize{5.81} \\
\bottomrule[1.5pt]
\end{tabular}}
\label{tab:app}
\end{center}
\end{table}

\begin{figure*}[!ht]
	\centering
	\includegraphics[width=0.95 \linewidth]{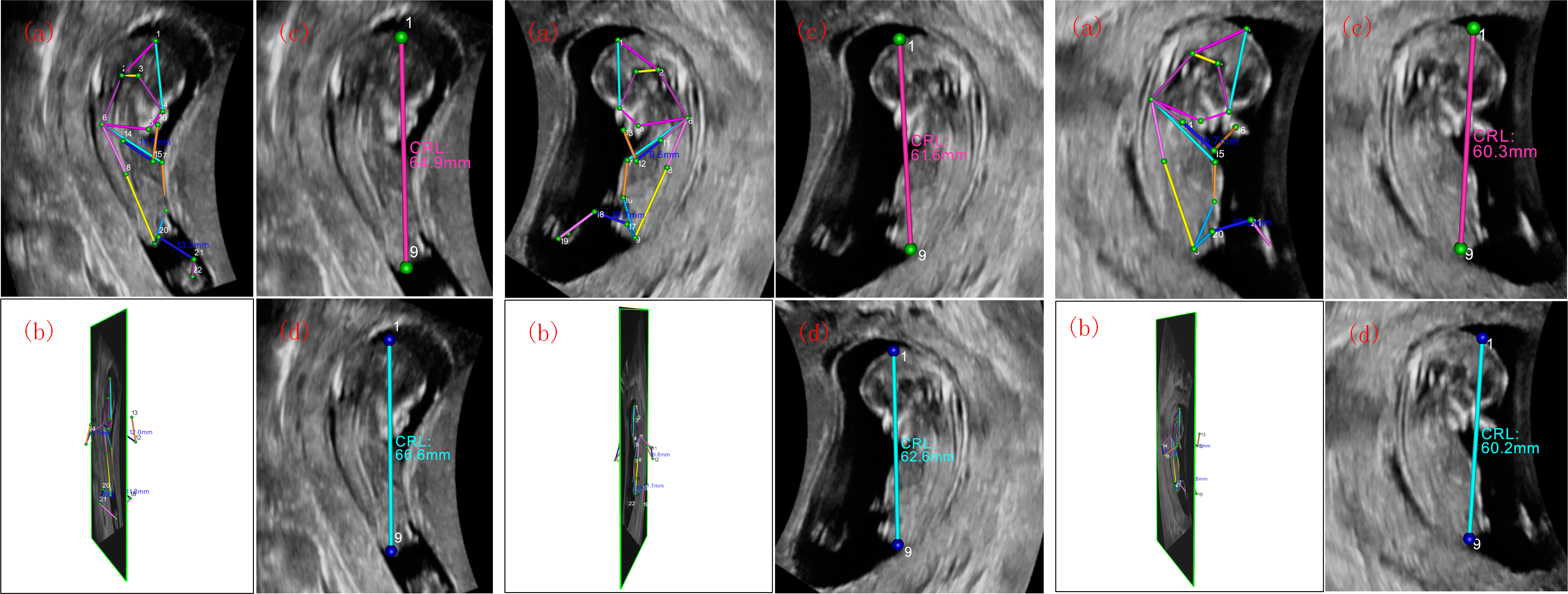}
	\caption{Clinical applications of \textbf{FetusMapV2}. (a) Lateral View. (b) Front View. (c) Prediction of Midsagittal Plane and CRL.(d) Ground Truth of Midsagittal Plane and CRL, where CRL is crown-rump length.}
	\label{fig:app_plane}
	\vspace{-0.3cm}	
\end{figure*}

\begin{figure*}[!ht]
	\centering
	\includegraphics[width=0.95 \linewidth]{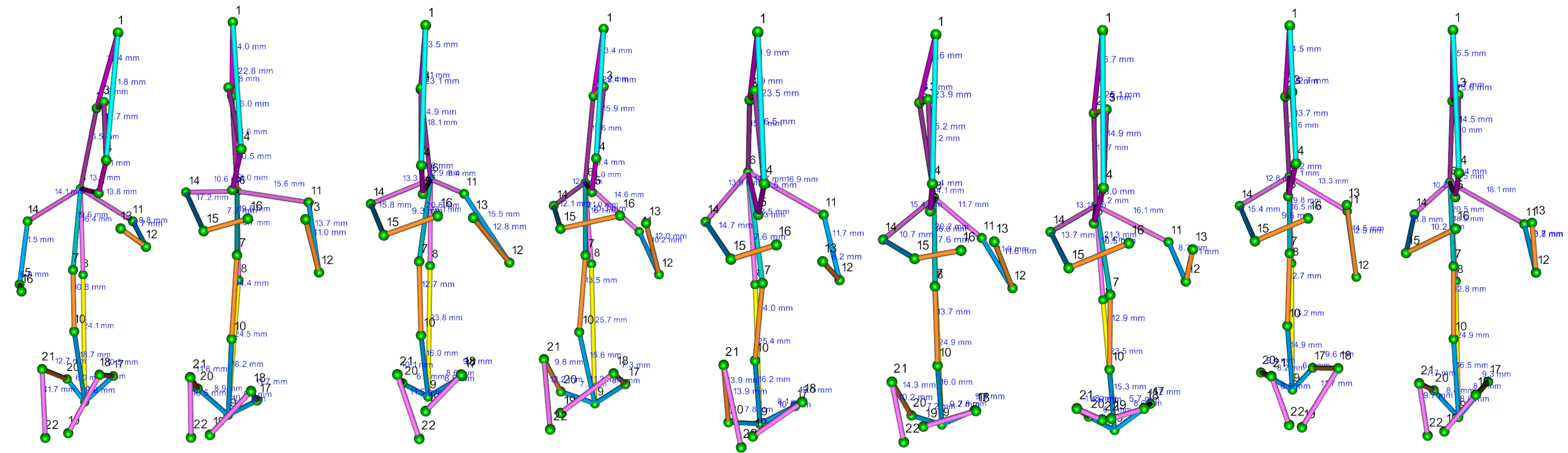}
	\caption{Application of fetal movement monitoring in 4D US images. }
	\label{fig:app_4d}	
\end{figure*}

\section{Applications}
Our framework can easily be extended to other clinically meaningful applications beyond fetal landmark detection. As shown in Table~\ref{tab:app}, we used the mean average error (MAE) to compute the difference between the prediction and ground truth to evaluate the effectiveness of our approach in the standard plane location and biometric measurement. For the midsagittal plane location, FetusMapV2 accurately locates several landmarks corresponding to anatomical structures, enabling the positional relationship between these landmarks to be used for the convenient and accurate locations of standard planes. Moreover, FetusMapV2 can also utilize the landmarks corresponding to the anatomical structures to accurately and efficiently measure key biometric parameters during fetal growth and development, such as long bone length (including the humerus and femur) and crown-rump length. The quantitative results in Table~\ref{tab:app} and the qualitative results in {{Fig.}}~\ref{fig:result} and ~\ref{fig:app_plane}  show that the results of FetusMapV2 in identifying the standing plane or measuring biometric parameters are more comparable to those annotated by experienced doctors. In addition, by generating a unique guide map for each fetus, FetusMapV2 allows the observer to locate specific anatomical parts quickly based on the information provided. Fig.~\ref{fig:app_4d} illustrates another exciting application of FetusMapV2 using 4D fetal ultrasound images to show the change in the pose of the fetus at different times, despite movement monitoring. 
Given the effectiveness of FetusMapV2 in localizing landmarks, we believe that it has the potential to be applied to other medical imaging tasks or to provide navigation for advanced studies.\par

\section{Discussion}
\label{discussion}
Our experimental results prove that increasing the resolution of the input can effectively improve the accuracy of landmark localization.
However, owing to the limited memory resources, it is difficult for common deep models to ensure that the input volume has a sufficiently high resolution.
Although several approaches can efficiently manage the GPU memory, they often consider only one type of technology (e.g., GCP, REV, etc.), thereby requiring heuristic models.

In this study, we formalized the resource management task as a graph partition problem to leverage different memory-management technologies for network architecture design simultaneously. 
Experimental results show that the effective combination of the advantages of different technologies can further reduce resource consumption and improve the model performance significantly while increasing only a little computation overhead during training. 
In future work, we will extend our proposed method to combine multiple (i.e., $>$2) advanced technologies more flexibly and effectively to further optimize resource utilization.

As shown in Tables~\ref{tab:all_methods_dis} and~\ref{tab:all_methods_auc}, the localization accuracy of the wrist and ankle landmarks(i.e., ${13, 16, 19, 22}$) was the worst among all other methods. Fig.~\ref{fig:Instance} (d)-(g) show several possible reasons. It can be observed that the wrist and ankle have the highest range of movement among all other landmarks. 
In addition, the anatomical characteristics of the two wrists and ankles are very similar, which confuses the network during the prediction of the two parts.
Although our proposed \emph{Pair Loss} alleviates the performance degradation caused by confusion to a certain extent, we believe that it cannot compensate for the severe lack of data with different poses. However, it is usually challenging to collect data covering all possible situations. Thus, improving the localization performance of these landmarks with limited training data is a future research topic for our study.

{Considering the poor quality of 3D fetal ultrasound volume, several anatomical structures may be 
indistinguishable from surrounding tissues (as shown in Fig.~\ref{fig:Instance}(b)). 
This poses challenges not only for the network to accurately locate these structures but also 
for experts to annotate them. Fortunately, we found that the relative relationship among the 
landmarks can be consider as a vital constraint, and it is reasonable to refine these ambiguous 
landmarks using distinguishable landmarks (i.e., have better localization performance). 
In this study, we observed that although the pose of the fetal limbs change significantly, 
the position between the head and trunk anatomical structures has relatively low degree of freedom. 
Therefore, we propose SSL framework to online refine these ambiguous landmarks, based on 
the robust shape prior knowledge. In addition, considering the diversity of fetal poses, 
big dataset is critical in advancing the pose estimation task. To the best of our knowledge, 
we have constructed the largest fetal ultrasound pose library in the literature ($S$, comprising 1000 cases), 
which encompasses a diverse range of fetal poses. Moreover, we are collaborating with more hospitals 
and recruiting volunteers at a much larger scale to further enrich our pose library.} \par
\section{Conclusion}
In this study, we propose an efficient, robust, and general framework for 3D fetal pose estimation. Our heuristic memory management framework combines the advantages of multiple recomputation technologies to manage training memory heuristically, producing a memory-efficient network architecture. 
The memory-efficient network can increase the input resolution to decrease landmark detection errors.
Moreover, to achieve robust results, we propose \emph{Pair Loss} to alleviate the confusion issues caused by symmetrical and similar landmarks.
We also introduced SSL based on the shape prior to remedy the relatively inadequate landmark localization results. It can improve the generalizability of the network.
The experiment results show that the proposed method achieves higher accuracy in landmark localization and pose estimation than other strong competitors. 
Based on the predicted key landmarks and the estimated pose, FetusMapV2 can not only directly benefit many clinical applications, but also guide numerous advanced studies.\par

\section{Acknowledge}
This work was supported by the grant from National Natural Science Foundation of China (Nos. 62171290, 62101343), Shenzhen-Hong Kong Joint Research Program (No. SGDX20201103095613036), Shenzhen Science and Technology Innovations Committee (No. 20200812143441001), and Hunan Provincial Natural Science Foundation of China (No. 2021JJ30173).

\bibliographystyle{model2-names.bst}\biboptions{authoryear}
\bibliography{refer}

\begin{thebibliography}{52}
\expandafter\ifx\csname natexlab\endcsname\relax\def\natexlab#1{#1}\fi
\providecommand{\url}[1]{\texttt{#1}}
\providecommand{\href}[2]{#2}
\providecommand{\path}[1]{#1}
\providecommand{\DOIprefix}{doi:}
\providecommand{\ArXivprefix}{arXiv:}
\providecommand{\URLprefix}{URL: }
\providecommand{\Pubmedprefix}{pmid:}
\providecommand{\doi}[1]{\href{http://dx.doi.org/#1}{\path{#1}}}
\providecommand{\Pubmed}[1]{\href{pmid:#1}{\path{#1}}}
\providecommand{\bibinfo}[2]{#2}
\ifx\xfnm\relax \def\xfnm[#1]{\unskip,\space#1}\fi
\bibitem[{Blumberg et~al.(2018)Blumberg, Tanno, Kokkinos and Alexander}]{blumberg2018deeper}
\bibinfo{author}{Blumberg, S.B.}, \bibinfo{author}{Tanno, R.}, \bibinfo{author}{Kokkinos, I.}, \bibinfo{author}{Alexander, D.C.}, \bibinfo{year}{2018}.
\newblock \bibinfo{title}{Deeper image quality transfer: Training low-memory neural networks for 3d images}, in: \bibinfo{booktitle}{MICCAI}, \bibinfo{organization}{Springer}. pp. \bibinfo{pages}{118--125}.
\bibitem[{Br{\"u}gger et~al.(2019)Br{\"u}gger, Baumgartner and Konukoglu}]{brugger2019partially}
\bibinfo{author}{Br{\"u}gger, R.}, \bibinfo{author}{Baumgartner, C.F.}, \bibinfo{author}{Konukoglu, E.}, \bibinfo{year}{2019}.
\newblock \bibinfo{title}{A partially reversible u-net for memory-efficient volumetric image segmentation}, in: \bibinfo{booktitle}{MICCAI}, \bibinfo{organization}{Springer}. pp. \bibinfo{pages}{429--437}.
\bibitem[{Cao et~al.(2018)Cao, Hidalgo, Simon, Wei and Sheikh}]{cao2018openpose}
\bibinfo{author}{Cao, Z.}, \bibinfo{author}{Hidalgo, G.}, \bibinfo{author}{Simon, T.}, \bibinfo{author}{Wei, S.E.}, \bibinfo{author}{Sheikh, Y.}, \bibinfo{year}{2018}.
\newblock \bibinfo{title}{Openpose: realtime multi-person 2d pose estimation using part affinity fields}.
\newblock \bibinfo{journal}{arXiv preprint arXiv:1812.08008} .
\bibitem[{Cao et~al.(2017)Cao, Simon, Wei and Sheikh}]{cv_pose_2d_2}
\bibinfo{author}{Cao, Z.}, \bibinfo{author}{Simon, T.}, \bibinfo{author}{Wei, S.E.}, \bibinfo{author}{Sheikh, Y.}, \bibinfo{year}{2017}.
\newblock \bibinfo{title}{Realtime multi-person 2d pose estimation using part affinity fields}, in: \bibinfo{booktitle}{Proceedings of the IEEE conference on computer vision and pattern recognition}, pp. \bibinfo{pages}{7291--7299}.
\bibitem[{Chen et~al.(2020)Chen, Yang, Huang, Shi, Liu, Lin, Huang, Yang, Zhang, Luo et~al.}]{chen2020region}
\bibinfo{author}{Chen, C.}, \bibinfo{author}{Yang, X.}, \bibinfo{author}{Huang, R.}, \bibinfo{author}{Shi, W.}, \bibinfo{author}{Liu, S.}, \bibinfo{author}{Lin, M.}, \bibinfo{author}{Huang, Y.}, \bibinfo{author}{Yang, Y.}, \bibinfo{author}{Zhang, Y.}, \bibinfo{author}{Luo, H.}, et~al., \bibinfo{year}{2020}.
\newblock \bibinfo{title}{Region proposal network with graph prior and iou-balance loss for landmark detection in 3d ultrasound}, in: \bibinfo{booktitle}{ISBI}, \bibinfo{organization}{IEEE}. pp. \bibinfo{pages}{1--5}.
\bibitem[{Chen et~al.(2016)Chen, Xu, Zhang and Guestrin}]{chen2016training}
\bibinfo{author}{Chen, T.}, \bibinfo{author}{Xu, B.}, \bibinfo{author}{Zhang, C.}, \bibinfo{author}{Guestrin, C.}, \bibinfo{year}{2016}.
\newblock \bibinfo{title}{Training deep nets with sublinear memory cost}.
\newblock \bibinfo{journal}{arXiv preprint arXiv:1604.06174} .
\bibitem[{Chen et~al.(2017)Chen, Shen, Wei, Liu and Yang}]{chen2017adversarial}
\bibinfo{author}{Chen, Y.}, \bibinfo{author}{Shen, C.}, \bibinfo{author}{Wei, X.S.}, \bibinfo{author}{Liu, L.}, \bibinfo{author}{Yang, J.}, \bibinfo{year}{2017}.
\newblock \bibinfo{title}{Adversarial posenet: A structure-aware convolutional network for human pose estimation}, in: \bibinfo{booktitle}{ICCV}, pp. \bibinfo{pages}{1212--1221}.
\bibitem[{Choi et~al.(2018)Choi, Lin, Xiang and Savarese}]{choi2018subcategory}
\bibinfo{author}{Choi, W.}, \bibinfo{author}{Lin, Y.}, \bibinfo{author}{Xiang, Y.}, \bibinfo{author}{Savarese, S.}, \bibinfo{year}{2018}.
\newblock \bibinfo{title}{Subcategory-aware convolutional neural networks for object detection}.
\newblock \bibinfo{note}{US Patent 9,965,719}.
\bibitem[{Christian~Etmann(2020)}]{Etmann2020iUNets}
\bibinfo{author}{Christian~Etmann, Rihuan~Ke, C.B.S.}, \bibinfo{year}{2020}.
\newblock \bibinfo{title}{iunets: Fully invertible u-nets with learnable up- and downsampling.}
\newblock \bibinfo{journal}{arXiv preprint} .
\bibitem[{{\c{C}}i{\c{c}}ek et~al.(2016){\c{C}}i{\c{c}}ek, Abdulkadir, Lienkamp, Brox and Ronneberger}]{cciccek20163d}
\bibinfo{author}{{\c{C}}i{\c{c}}ek, {\"O}.}, \bibinfo{author}{Abdulkadir, A.}, \bibinfo{author}{Lienkamp, S.S.}, \bibinfo{author}{Brox, T.}, \bibinfo{author}{Ronneberger, O.}, \bibinfo{year}{2016}.
\newblock \bibinfo{title}{3d u-net: learning dense volumetric segmentation from sparse annotation}, in: \bibinfo{booktitle}{MICCAI}, \bibinfo{organization}{Springer}. pp. \bibinfo{pages}{424--432}.
\bibitem[{Fang et~al.(2022)Fang, Li, Tang, Xu, Zhu, Xiu, Li and Lu}]{action_reg}
\bibinfo{author}{Fang, H.S.}, \bibinfo{author}{Li, J.}, \bibinfo{author}{Tang, H.}, \bibinfo{author}{Xu, C.}, \bibinfo{author}{Zhu, H.}, \bibinfo{author}{Xiu, Y.}, \bibinfo{author}{Li, Y.L.}, \bibinfo{author}{Lu, C.}, \bibinfo{year}{2022}.
\newblock \bibinfo{title}{Alphapose: Whole-body regional multi-person pose estimation and tracking in real-time}.
\newblock \bibinfo{journal}{IEEE Transactions on Pattern Analysis and Machine Intelligence} .
\bibitem[{Golan et~al.(2016)Golan, Donner, Mansi, Jaremko, Ramachandran et~al.}]{golan2016fully}
\bibinfo{author}{Golan, D.}, \bibinfo{author}{Donner, Y.}, \bibinfo{author}{Mansi, C.}, \bibinfo{author}{Jaremko, J.}, \bibinfo{author}{Ramachandran, M.}, et~al., \bibinfo{year}{2016}.
\newblock \bibinfo{title}{Fully automating graf’s method for ddh diagnosis using deep convolutional neural networks}, in: \bibinfo{booktitle}{Deep Learning and Data Labeling for Medical Applications}. \bibinfo{publisher}{Springer}, pp. \bibinfo{pages}{130--141}.
\bibitem[{Gomez et~al.(2017)Gomez, Ren, Urtasun and Grosse}]{gomez2017reversible}
\bibinfo{author}{Gomez, A.N.}, \bibinfo{author}{Ren, M.}, \bibinfo{author}{Urtasun, R.}, \bibinfo{author}{Grosse, R.B.}, \bibinfo{year}{2017}.
\newblock \bibinfo{title}{The reversible residual network: Backpropagation without storing activations}, in: \bibinfo{booktitle}{NIPS}, pp. \bibinfo{pages}{2214--2224}.
\bibitem[{Grandjean et~al.(2018)Grandjean, Hossu, Bertholdt, Noble, Morel and Grang{\'e}}]{fetal_3d_measurement}
\bibinfo{author}{Grandjean, G.A.}, \bibinfo{author}{Hossu, G.}, \bibinfo{author}{Bertholdt, C.}, \bibinfo{author}{Noble, P.}, \bibinfo{author}{Morel, O.}, \bibinfo{author}{Grang{\'e}, G.}, \bibinfo{year}{2018}.
\newblock \bibinfo{title}{Artificial intelligence assistance for fetal head biometry: assessment of automated measurement software}.
\newblock \bibinfo{journal}{Diagnostic and interventional imaging} \bibinfo{volume}{99}, \bibinfo{pages}{709--716}.
\bibitem[{Gruslys et~al.(2016)Gruslys, Munos, Danihelka, Lanctot and Graves}]{gruslys2016memory}
\bibinfo{author}{Gruslys, A.}, \bibinfo{author}{Munos, R.}, \bibinfo{author}{Danihelka, I.}, \bibinfo{author}{Lanctot, M.}, \bibinfo{author}{Graves, A.}, \bibinfo{year}{2016}.
\newblock \bibinfo{title}{Memory-efficient backpropagation through time}, in: \bibinfo{booktitle}{NIPS}, pp. \bibinfo{pages}{4125--4133}.
\bibitem[{Guo and Yuan(2019)}]{guo2019triple}
\bibinfo{author}{Guo, X.}, \bibinfo{author}{Yuan, Y.}, \bibinfo{year}{2019}.
\newblock \bibinfo{title}{Triple anet: Adaptive abnormal-aware attention network for wce image classification}, in: \bibinfo{booktitle}{MICCAI}, \bibinfo{organization}{Springer}. pp. \bibinfo{pages}{293--301}.
\bibitem[{Hu et~al.(2021)Hu, Wang, Yang, Zhou, Xue, Cao, Liu, Huang, Guo, Shang et~al.}]{hu2021joint}
\bibinfo{author}{Hu, X.}, \bibinfo{author}{Wang, L.}, \bibinfo{author}{Yang, X.}, \bibinfo{author}{Zhou, X.}, \bibinfo{author}{Xue, W.}, \bibinfo{author}{Cao, Y.}, \bibinfo{author}{Liu, S.}, \bibinfo{author}{Huang, Y.}, \bibinfo{author}{Guo, S.}, \bibinfo{author}{Shang, N.}, et~al., \bibinfo{year}{2021}.
\newblock \bibinfo{title}{Joint landmark and structure learning for automatic evaluation of developmental dysplasia of the hip}.
\newblock \bibinfo{journal}{IEEE Journal of Biomedical and Health Informatics} \bibinfo{volume}{26}, \bibinfo{pages}{345--358}.
\bibitem[{Huang et~al.(2018)Huang, Noble and Namburete}]{huang2018omni}
\bibinfo{author}{Huang, R.}, \bibinfo{author}{Noble, J.A.}, \bibinfo{author}{Namburete, A.I.}, \bibinfo{year}{2018}.
\newblock \bibinfo{title}{Omni-supervised learning: scaling up to large unlabelled medical datasets}, in: \bibinfo{booktitle}{MICCAI}, \bibinfo{organization}{Springer}. pp. \bibinfo{pages}{572--580}.
\bibitem[{Huang et~al.(2021)Huang, Yang, Zou, Chen, Wang, Dou, Ravikumar, Frangi, Zhou and Ni}]{huang2021flip}
\bibinfo{author}{Huang, Y.}, \bibinfo{author}{Yang, X.}, \bibinfo{author}{Zou, Y.}, \bibinfo{author}{Chen, C.}, \bibinfo{author}{Wang, J.}, \bibinfo{author}{Dou, H.}, \bibinfo{author}{Ravikumar, N.}, \bibinfo{author}{Frangi, A.F.}, \bibinfo{author}{Zhou, J.}, \bibinfo{author}{Ni, D.}, \bibinfo{year}{2021}.
\newblock \bibinfo{title}{Flip learning: Erase to segment}, in: \bibinfo{booktitle}{International Conference on Medical Image Computing and Computer-Assisted Intervention}, \bibinfo{organization}{Springer}. pp. \bibinfo{pages}{493--502}.
\bibitem[{Isensee et~al.(2018)Isensee, Kickingereder, Wick, Bendszus and Maier-Hein}]{isensee2018no}
\bibinfo{author}{Isensee, F.}, \bibinfo{author}{Kickingereder, P.}, \bibinfo{author}{Wick, W.}, \bibinfo{author}{Bendszus, M.}, \bibinfo{author}{Maier-Hein, K.H.}, \bibinfo{year}{2018}.
\newblock \bibinfo{title}{No new-net}, in: \bibinfo{booktitle}{MICCAI}, \bibinfo{organization}{Springer}. pp. \bibinfo{pages}{234--244}.
\bibitem[{Jang et~al.(2017)Jang, Kwon, Kim, Lee, Park and Seo}]{jang2017cnn}
\bibinfo{author}{Jang, J.}, \bibinfo{author}{Kwon, J.Y.}, \bibinfo{author}{Kim, B.}, \bibinfo{author}{Lee, S.M.}, \bibinfo{author}{Park, Y.}, \bibinfo{author}{Seo, J.K.}, \bibinfo{year}{2017}.
\newblock \bibinfo{title}{Cnn-based estimation of abdominal circumference from ultrasound images.}
\newblock \bibinfo{journal}{arXiv preprint arXiv:1702.02741} .
\bibitem[{Khawam et~al.(2021)Khawam, De~Dumast, Deman, Kebiri, Yu, Tourbier, Lajous, Hagmann, Maeder, Thiran et~al.}]{fetal_3d_measurement_2}
\bibinfo{author}{Khawam, M.}, \bibinfo{author}{De~Dumast, P.}, \bibinfo{author}{Deman, P.}, \bibinfo{author}{Kebiri, H.}, \bibinfo{author}{Yu, T.}, \bibinfo{author}{Tourbier, S.}, \bibinfo{author}{Lajous, H.}, \bibinfo{author}{Hagmann, P.}, \bibinfo{author}{Maeder, P.}, \bibinfo{author}{Thiran, J.P.}, et~al., \bibinfo{year}{2021}.
\newblock \bibinfo{title}{Fetal brain biometric measurements on 3d super-resolution reconstructed t2-weighted mri: An intra-and inter-observer agreement study}.
\newblock \bibinfo{journal}{Frontiers in Pediatrics} \bibinfo{volume}{9}, \bibinfo{pages}{639746}.
\bibitem[{Kingma and Ba(2014)}]{kingma2014adam}
\bibinfo{author}{Kingma, D.P.}, \bibinfo{author}{Ba, J.}, \bibinfo{year}{2014}.
\newblock \bibinfo{title}{Adam: A method for stochastic optimization}.
\newblock \bibinfo{journal}{arXiv preprint arXiv:1412.6980} .
\bibitem[{Kumar et~al.(2019)Kumar, Purohit, Svitkina, Vee and Wang}]{kumar2019efficient}
\bibinfo{author}{Kumar, R.}, \bibinfo{author}{Purohit, M.}, \bibinfo{author}{Svitkina, Z.}, \bibinfo{author}{Vee, E.}, \bibinfo{author}{Wang, J.}, \bibinfo{year}{2019}.
\newblock \bibinfo{title}{Efficient rematerialization for deep networks}, in: \bibinfo{booktitle}{NIPS}, pp. \bibinfo{pages}{15146--15155}.
\bibitem[{Kusumoto et~al.(2019)Kusumoto, Inoue, Watanabe, Akiba and Koyama}]{kusumoto2019graph}
\bibinfo{author}{Kusumoto, M.}, \bibinfo{author}{Inoue, T.}, \bibinfo{author}{Watanabe, G.}, \bibinfo{author}{Akiba, T.}, \bibinfo{author}{Koyama, M.}, \bibinfo{year}{2019}.
\newblock \bibinfo{title}{A graph theoretic framework of recomputation algorithms for memory-efficient backpropagation}, in: \bibinfo{booktitle}{NIPS}, pp. \bibinfo{pages}{1161--1170}.
\bibitem[{Li et~al.(2018a)Li, Khanal, Hou, Alansary, Cerrolaza, Sinclair, Matthew, Gupta, Knight, Kainz et~al.}]{fetal_plane_3d}
\bibinfo{author}{Li, Y.}, \bibinfo{author}{Khanal, B.}, \bibinfo{author}{Hou, B.}, \bibinfo{author}{Alansary, A.}, \bibinfo{author}{Cerrolaza, J.J.}, \bibinfo{author}{Sinclair, M.}, \bibinfo{author}{Matthew, J.}, \bibinfo{author}{Gupta, C.}, \bibinfo{author}{Knight, C.}, \bibinfo{author}{Kainz, B.}, et~al., \bibinfo{year}{2018}a.
\newblock \bibinfo{title}{Standard plane detection in 3d fetal ultrasound using an iterative transformation network}, in: \bibinfo{booktitle}{Medical Image Computing and Computer Assisted Intervention--MICCAI 2018: 21st International Conference, Granada, Spain, September 16-20, 2018, Proceedings, Part I}, \bibinfo{organization}{Springer}. pp. \bibinfo{pages}{392--400}.
\bibitem[{Li et~al.(2018b)Li, Wang, Ji, Xiang and Fox}]{motion_cap}
\bibinfo{author}{Li, Y.}, \bibinfo{author}{Wang, G.}, \bibinfo{author}{Ji, X.}, \bibinfo{author}{Xiang, Y.}, \bibinfo{author}{Fox, D.}, \bibinfo{year}{2018}b.
\newblock \bibinfo{title}{Deepim: Deep iterative matching for 6d pose estimation}, in: \bibinfo{booktitle}{Proceedings of the European Conference on Computer Vision (ECCV)}, pp. \bibinfo{pages}{683--698}.
\bibitem[{Liang et~al.(2022)Liang, Yang, Huang, Li, He, Hu, Chen, Xue, Cheng and Ni}]{liang2022sketch}
\bibinfo{author}{Liang, J.}, \bibinfo{author}{Yang, X.}, \bibinfo{author}{Huang, Y.}, \bibinfo{author}{Li, H.}, \bibinfo{author}{He, S.}, \bibinfo{author}{Hu, X.}, \bibinfo{author}{Chen, Z.}, \bibinfo{author}{Xue, W.}, \bibinfo{author}{Cheng, J.}, \bibinfo{author}{Ni, D.}, \bibinfo{year}{2022}.
\newblock \bibinfo{title}{Sketch guided and progressive growing gan for realistic and editable ultrasound image synthesis}.
\newblock \bibinfo{journal}{Medical Image Analysis} \bibinfo{volume}{79}, \bibinfo{pages}{102461}.
\bibitem[{Liu et~al.(2019)Liu, Ding, Shahroudy, Duan, Jiang, Wang and Chichung}]{liu2019feature}
\bibinfo{author}{Liu, J.}, \bibinfo{author}{Ding, H.}, \bibinfo{author}{Shahroudy, A.}, \bibinfo{author}{Duan, L.Y.}, \bibinfo{author}{Jiang, X.}, \bibinfo{author}{Wang, G.}, \bibinfo{author}{Chichung, A.K.}, \bibinfo{year}{2019}.
\newblock \bibinfo{title}{Feature boosting network for 3d pose estimation}.
\newblock \bibinfo{journal}{PAMI} .
\bibitem[{Martinez et~al.(2017)Martinez, Hossain, Romero and Little}]{cv_pose_3d}
\bibinfo{author}{Martinez, J.}, \bibinfo{author}{Hossain, R.}, \bibinfo{author}{Romero, J.}, \bibinfo{author}{Little, J.J.}, \bibinfo{year}{2017}.
\newblock \bibinfo{title}{A simple yet effective baseline for 3d human pose estimation}, in: \bibinfo{booktitle}{Proceedings of the IEEE international conference on computer vision}, pp. \bibinfo{pages}{2640--2649}.
\bibitem[{Micikevicius et~al.(2017)Micikevicius, Narang, Alben, Diamos, Elsen, Garcia, Ginsburg, Houston, Kuchaiev, Venkatesh et~al.}]{micikevicius2017mixed}
\bibinfo{author}{Micikevicius, P.}, \bibinfo{author}{Narang, S.}, \bibinfo{author}{Alben, J.}, \bibinfo{author}{Diamos, G.}, \bibinfo{author}{Elsen, E.}, \bibinfo{author}{Garcia, D.}, \bibinfo{author}{Ginsburg, B.}, \bibinfo{author}{Houston, M.}, \bibinfo{author}{Kuchaiev, O.}, \bibinfo{author}{Venkatesh, G.}, et~al., \bibinfo{year}{2017}.
\newblock \bibinfo{title}{Mixed precision training}.
\newblock \bibinfo{journal}{arXiv preprint arXiv:1710.03740} .
\bibitem[{Mostafa and Wang(2019)}]{mostafa2019parameter}
\bibinfo{author}{Mostafa, H.}, \bibinfo{author}{Wang, X.}, \bibinfo{year}{2019}.
\newblock \bibinfo{title}{Parameter efficient training of deep convolutional neural networks by dynamic sparse reparameterization}.
\newblock \bibinfo{journal}{arXiv preprint arXiv:1902.05967} .
\bibitem[{Napolitano et~al.(2020)Napolitano, Molloholli, Donadono, Ohuma, Wanyonyi, Kemp, Yaqub, Ash, Barros, Carvalho et~al.}]{alison_measure}
\bibinfo{author}{Napolitano, R.}, \bibinfo{author}{Molloholli, M.}, \bibinfo{author}{Donadono, V.}, \bibinfo{author}{Ohuma, E.O.}, \bibinfo{author}{Wanyonyi, S.}, \bibinfo{author}{Kemp, B.}, \bibinfo{author}{Yaqub, M.K.}, \bibinfo{author}{Ash, S.}, \bibinfo{author}{Barros, F.C.}, \bibinfo{author}{Carvalho, M.}, et~al., \bibinfo{year}{2020}.
\newblock \bibinfo{title}{International standards for fetal brain structures based on serial ultrasound measurements from fetal growth longitudinal study of intergrowth-21st project}.
\newblock \bibinfo{journal}{Ultrasound in Obstetrics \& Gynecology} \bibinfo{volume}{56}, \bibinfo{pages}{359--370}.
\bibitem[{Newell et~al.(2016a)Newell, Yang and Deng}]{cvpose_2d}
\bibinfo{author}{Newell, A.}, \bibinfo{author}{Yang, K.}, \bibinfo{author}{Deng, J.}, \bibinfo{year}{2016}a.
\newblock \bibinfo{title}{Stacked hourglass networks for human pose estimation}, in: \bibinfo{booktitle}{Computer Vision--ECCV 2016: 14th European Conference, Amsterdam, The Netherlands, October 11-14, 2016, Proceedings, Part VIII 14}, \bibinfo{organization}{Springer}. pp. \bibinfo{pages}{483--499}.
\bibitem[{Newell et~al.(2016b)Newell, Yang and Deng}]{newell2016stacked}
\bibinfo{author}{Newell, A.}, \bibinfo{author}{Yang, K.}, \bibinfo{author}{Deng, J.}, \bibinfo{year}{2016}b.
\newblock \bibinfo{title}{Stacked hourglass networks for human pose estimation}, in: \bibinfo{booktitle}{ECCV}, \bibinfo{organization}{Springer}. pp. \bibinfo{pages}{483--499}.
\bibitem[{Ni et~al.(2023)Ni, Xue, Ma, Zhang, Li and Huang}]{NI2023102654}
\bibinfo{author}{Ni, H.}, \bibinfo{author}{Xue, Y.}, \bibinfo{author}{Ma, L.}, \bibinfo{author}{Zhang, Q.}, \bibinfo{author}{Li, X.}, \bibinfo{author}{Huang, S.X.}, \bibinfo{year}{2023}.
\newblock \bibinfo{title}{Semi-supervised body parsing and pose estimation for enhancing infant general movement assessment}.
\newblock \bibinfo{journal}{Medical Image Analysis} \bibinfo{volume}{83}, \bibinfo{pages}{102654}.
\bibitem[{Noble and Boukerroui(2006)}]{noble2006ultrasound}
\bibinfo{author}{Noble, J.A.}, \bibinfo{author}{Boukerroui, D.}, \bibinfo{year}{2006}.
\newblock \bibinfo{title}{Ultrasound image segmentation: a survey}.
\newblock \bibinfo{journal}{TMI} \bibinfo{volume}{25}, \bibinfo{pages}{987--1010}.
\bibitem[{Paszke et~al.(2019)Paszke, Gross, Massa, Lerer, Bradbury, Chanan, Killeen, Lin, Gimelshein, Antiga et~al.}]{paszke2019pytorch}
\bibinfo{author}{Paszke, A.}, \bibinfo{author}{Gross, S.}, \bibinfo{author}{Massa, F.}, \bibinfo{author}{Lerer, A.}, \bibinfo{author}{Bradbury, J.}, \bibinfo{author}{Chanan, G.}, \bibinfo{author}{Killeen, T.}, \bibinfo{author}{Lin, Z.}, \bibinfo{author}{Gimelshein, N.}, \bibinfo{author}{Antiga, L.}, et~al., \bibinfo{year}{2019}.
\newblock \bibinfo{title}{Pytorch: An imperative style, high-performance deep learning library}, in: \bibinfo{booktitle}{NIPS}, pp. \bibinfo{pages}{8024--8035}.
\bibitem[{Payer et~al.(2016)Payer, {\v{S}}tern, Bischof and Urschler}]{payer2016regressing}
\bibinfo{author}{Payer, C.}, \bibinfo{author}{{\v{S}}tern, D.}, \bibinfo{author}{Bischof, H.}, \bibinfo{author}{Urschler, M.}, \bibinfo{year}{2016}.
\newblock \bibinfo{title}{Regressing heatmaps for multiple landmark localization using cnns}, in: \bibinfo{booktitle}{MICCAI}, \bibinfo{organization}{Springer}. pp. \bibinfo{pages}{230--238}.
\bibitem[{Pleiss et~al.(2017)Pleiss, Chen, Huang, Li, van~der Maaten and Weinberger}]{pleiss2017memory}
\bibinfo{author}{Pleiss, G.}, \bibinfo{author}{Chen, D.}, \bibinfo{author}{Huang, G.}, \bibinfo{author}{Li, T.}, \bibinfo{author}{van~der Maaten, L.}, \bibinfo{author}{Weinberger, K.Q.}, \bibinfo{year}{2017}.
\newblock \bibinfo{title}{Memory-efficient implementation of densenets}.
\newblock \bibinfo{journal}{arXiv preprint arXiv:1707.06990} .
\bibitem[{Reddy et~al.(2008)Reddy, Filly and Copel}]{reddy2008prenatal}
\bibinfo{author}{Reddy, U.M.}, \bibinfo{author}{Filly, R.A.}, \bibinfo{author}{Copel, J.A.}, \bibinfo{year}{2008}.
\newblock \bibinfo{title}{Prenatal imaging: ultrasonography and magnetic resonance imaging}.
\newblock \bibinfo{journal}{Obstetrics and gynecology} \bibinfo{volume}{112}, \bibinfo{pages}{145}.
\bibitem[{Sofka et~al.(2017)Sofka, Milletari, Jia and Rothberg}]{sofka2017fully}
\bibinfo{author}{Sofka, M.}, \bibinfo{author}{Milletari, F.}, \bibinfo{author}{Jia, J.}, \bibinfo{author}{Rothberg, A.}, \bibinfo{year}{2017}.
\newblock \bibinfo{title}{Fully convolutional regression network for accurate detection of measurement points}, in: \bibinfo{booktitle}{DLMIA}. \bibinfo{publisher}{Springer}, pp. \bibinfo{pages}{258--266}.
\bibitem[{Wang et~al.(2020)Wang, Miao, Yang, Li, Zhou, Huang, Lin, Xue, Jia, Zhou et~al.}]{wang2020auto}
\bibinfo{author}{Wang, J.}, \bibinfo{author}{Miao, J.}, \bibinfo{author}{Yang, X.}, \bibinfo{author}{Li, R.}, \bibinfo{author}{Zhou, G.}, \bibinfo{author}{Huang, Y.}, \bibinfo{author}{Lin, Z.}, \bibinfo{author}{Xue, W.}, \bibinfo{author}{Jia, X.}, \bibinfo{author}{Zhou, J.}, et~al., \bibinfo{year}{2020}.
\newblock \bibinfo{title}{Auto-weighting for breast cancer classification in multimodal ultrasound}, in: \bibinfo{booktitle}{International Conference on Medical Image Computing and Computer-Assisted Intervention}, \bibinfo{organization}{Springer}. pp. \bibinfo{pages}{190--199}.
\bibitem[{Wang et~al.(2018)Wang, Zhao and Ji}]{human_computer_interaction}
\bibinfo{author}{Wang, K.}, \bibinfo{author}{Zhao, R.}, \bibinfo{author}{Ji, Q.}, \bibinfo{year}{2018}.
\newblock \bibinfo{title}{Human computer interaction with head pose, eye gaze and body gestures}, in: \bibinfo{booktitle}{2018 13th IEEE International Conference on Automatic Face \& Gesture Recognition (FG 2018)}, \bibinfo{organization}{IEEE}. pp. \bibinfo{pages}{789--789}.
\bibitem[{Xu et~al.(2020)Xu, Zhang, Turk, Grant, Golland and Adalsteinsson}]{xu20203d}
\bibinfo{author}{Xu, J.}, \bibinfo{author}{Zhang, M.}, \bibinfo{author}{Turk, E.A.}, \bibinfo{author}{Grant, P.E.}, \bibinfo{author}{Golland, P.}, \bibinfo{author}{Adalsteinsson, E.}, \bibinfo{year}{2020}.
\newblock \bibinfo{title}{3d fetal pose estimation with adaptive variance and conditional generative adversarial network}, in: \bibinfo{booktitle}{Medical Ultrasound, and Preterm, Perinatal and Paediatric Image Analysis}. \bibinfo{publisher}{Springer}, pp. \bibinfo{pages}{201--210}.
\bibitem[{Xu et~al.(2019)Xu, Zhang, Turk, Zhang, Grant, Ying, Golland and Adalsteinsson}]{xu2019fetal}
\bibinfo{author}{Xu, J.}, \bibinfo{author}{Zhang, M.}, \bibinfo{author}{Turk, E.A.}, \bibinfo{author}{Zhang, L.}, \bibinfo{author}{Grant, P.E.}, \bibinfo{author}{Ying, K.}, \bibinfo{author}{Golland, P.}, \bibinfo{author}{Adalsteinsson, E.}, \bibinfo{year}{2019}.
\newblock \bibinfo{title}{Fetal pose estimation in volumetric mri using a 3d convolution neural network}, in: \bibinfo{booktitle}{MICCAI}, \bibinfo{organization}{Springer}. pp. \bibinfo{pages}{403--410}.
\bibitem[{Xu et~al.(2018)Xu, Huo, Park, Landman, Milkowski, Grbic and Zhou}]{xu2018less}
\bibinfo{author}{Xu, Z.}, \bibinfo{author}{Huo, Y.}, \bibinfo{author}{Park, J.}, \bibinfo{author}{Landman, B.}, \bibinfo{author}{Milkowski, A.}, \bibinfo{author}{Grbic, S.}, \bibinfo{author}{Zhou, S.}, \bibinfo{year}{2018}.
\newblock \bibinfo{title}{Less is more: Simultaneous view classification and landmark detection for abdominal ultrasound images}, in: \bibinfo{booktitle}{MICCAI}, \bibinfo{organization}{Springer}. pp. \bibinfo{pages}{711--719}.
\bibitem[{Yang et~al.(2021a)Yang, Dou, Huang, Xue, Huang, Qian, Zhang, Luo, Guo, Wang et~al.}]{yang2021agent}
\bibinfo{author}{Yang, X.}, \bibinfo{author}{Dou, H.}, \bibinfo{author}{Huang, R.}, \bibinfo{author}{Xue, W.}, \bibinfo{author}{Huang, Y.}, \bibinfo{author}{Qian, J.}, \bibinfo{author}{Zhang, Y.}, \bibinfo{author}{Luo, H.}, \bibinfo{author}{Guo, H.}, \bibinfo{author}{Wang, T.}, et~al., \bibinfo{year}{2021}a.
\newblock \bibinfo{title}{Agent with warm start and adaptive dynamic termination for plane localization in 3d ultrasound}.
\newblock \bibinfo{journal}{IEEE Transactions on Medical Imaging} \bibinfo{volume}{40}, \bibinfo{pages}{1950--1961}.
\bibitem[{Yang et~al.(2021b)Yang, Huang, Huang, Dou, Li, Qian, Huang, Shi, Chen, Zhang et~al.}]{yang2021searching}
\bibinfo{author}{Yang, X.}, \bibinfo{author}{Huang, Y.}, \bibinfo{author}{Huang, R.}, \bibinfo{author}{Dou, H.}, \bibinfo{author}{Li, R.}, \bibinfo{author}{Qian, J.}, \bibinfo{author}{Huang, X.}, \bibinfo{author}{Shi, W.}, \bibinfo{author}{Chen, C.}, \bibinfo{author}{Zhang, Y.}, et~al., \bibinfo{year}{2021}b.
\newblock \bibinfo{title}{Searching collaborative agents for multi-plane localization in 3d ultrasound}.
\newblock \bibinfo{journal}{Medical Image Analysis} \bibinfo{volume}{72}, \bibinfo{pages}{102119}.
\bibitem[{Yang et~al.(2019)Yang, Shi, Dou, Qian, Wang, Xue, Li, Ni and Heng}]{yang2019fetusmap}
\bibinfo{author}{Yang, X.}, \bibinfo{author}{Shi, W.}, \bibinfo{author}{Dou, H.}, \bibinfo{author}{Qian, J.}, \bibinfo{author}{Wang, Y.}, \bibinfo{author}{Xue, W.}, \bibinfo{author}{Li, S.}, \bibinfo{author}{Ni, D.}, \bibinfo{author}{Heng, P.A.}, \bibinfo{year}{2019}.
\newblock \bibinfo{title}{Fetusmap: Fetal pose estimation in 3d ultrasound}, in: \bibinfo{booktitle}{MICCAI}, \bibinfo{organization}{Springer}. pp. \bibinfo{pages}{281--289}.
\bibitem[{Yang et~al.(2017)Yang, Yu, Li, Wang, Wang, Qin, Ni and Heng}]{yang2017towards}
\bibinfo{author}{Yang, X.}, \bibinfo{author}{Yu, L.}, \bibinfo{author}{Li, S.}, \bibinfo{author}{Wang, X.}, \bibinfo{author}{Wang, N.}, \bibinfo{author}{Qin, J.}, \bibinfo{author}{Ni, D.}, \bibinfo{author}{Heng, P.A.}, \bibinfo{year}{2017}.
\newblock \bibinfo{title}{Towards automatic semantic segmentation in volumetric ultrasound}, in: \bibinfo{booktitle}{MICCAI}, \bibinfo{organization}{Springer}. pp. \bibinfo{pages}{711--719}.
\bibitem[{Zonoobi et~al.(2018)Zonoobi, Hareendranathan, Mostofi, Mabee, Pasha, Cobzas, Rao, Dulai, Kapur and Jaremko}]{zonoobi2018developmental}
\bibinfo{author}{Zonoobi, D.}, \bibinfo{author}{Hareendranathan, A.}, \bibinfo{author}{Mostofi, E.}, \bibinfo{author}{Mabee, M.}, \bibinfo{author}{Pasha, S.}, \bibinfo{author}{Cobzas, D.}, \bibinfo{author}{Rao, P.}, \bibinfo{author}{Dulai, S.K.}, \bibinfo{author}{Kapur, J.}, \bibinfo{author}{Jaremko, J.L.}, \bibinfo{year}{2018}.
\newblock \bibinfo{title}{Developmental hip dysplasia diagnosis at three-dimensional us: a multicenter study}.
\newblock \bibinfo{journal}{Radiology} \bibinfo{volume}{287}, \bibinfo{pages}{1003--1015}.

\end{thebibliography}

\end{document}